\documentclass[aps,prd,superscriptaddress,preprint,secnumarabic,nofootinbib]{revtex4-2}
\pdfoutput=1
\usepackage[utf8]{inputenc}
\usepackage[normalem]{ulem} 
\usepackage[english]{babel}
\usepackage{tabularx}
\usepackage{array}
\usepackage{graphics}
\usepackage[pdftex]{graphicx}
\usepackage{mathtools}
\usepackage{amssymb}
\usepackage{amsmath}
\usepackage{slashed}
\usepackage{braket}
\usepackage{bm}

\usepackage[usenames,dvipsnames]{xcolor}
\usepackage{bm}

\newcommand {\e}[1]{\mathrm{\,#1}}       
\newcommand{\mc}[1]{\mathcal{#1}}
\newcommand{\mrm}[1]{\mathrm{#1}}

\newcommand{\re}[0]{\mrm{Re}}

\definecolor{niceblue}{rgb}{0.15,0.15,0.6}
\definecolor{nicegreen}{rgb}{0.1,0.5,0.1}
\definecolor{Red}{rgb}{1.,0.,0.}

\definecolor{Green}{rgb}{0.2,.7,0.2}

\definecolor{nicered}{rgb}{0.7,0.1,0.1}
\definecolor{nicegreen}{rgb}{0.1,0.5,0.1}
\definecolor{violet}{rgb}{0.7,0.3,0.3}

\newcommand{\rddst}{R_{D^{(\ast)}}}
\newcommand{\Li}[0]{\mathrm{Li}_2}

\begin{document}

\title{$\rddst$ and survival of the fittest scalar leptoquark}

\author{Damir Bečirević} \email[Electronic address:]{\ damir.becirevic@ijclab.in2p3.fr}
\affiliation{IJCLab, P\^ole Th\'eorie (Bat.~210), CNRS/IN2P3 and \\ 
Universit\'e, Paris-Saclay, 91405 Orsay, France}

\author{Svjetlana Fajfer} \email[Electronic address:]{\ svjetlana.fajfer@ijs.si}
\affiliation{Department of Physics, University of Ljubljana, Jadranska 19, 1000 Ljubljana, Slovenia}
\affiliation{Jo\v zef Stefan Institute, Jamova 39, P.\ O.\ Box 3000, 1001
  Ljubljana, Slovenia}

\author{Nejc Košnik} \email[Electronic address:]{\ nejc.kosnik@ijs.si}
\affiliation{Department of Physics, University of Ljubljana, Jadranska 19, 1000 Ljubljana, Slovenia}
\affiliation{Jo\v zef Stefan Institute, Jamova 39, P.\ O.\ Box 3000, 1001
  Ljubljana, Slovenia}

\author{Lovre Pavičić} \email[Electronic address:]{\ lovre.pavicic@ijs.si}
\affiliation{Jo\v zef Stefan Institute, Jamova 39, P.\ O.\ Box 3000, 1001
  Ljubljana, Slovenia}

\begin{abstract}
Motivated by the long-standing discrepancy in lepton flavor universality ratios $R_D$ and $R_{D^{\ast}}$ we assess the status of scalar leptoquark states $R_2$, $\widetilde R_2$ and $S_1$ which can in principle provide a desired enhancement of $\mathcal{B}(B\to D^{(\ast )}\tau \nu)$ in a minimal setup with two Yukawa couplings only. We consider unavoidable low-energy constraints, $Z$-pole measurements as well as high-$p_T$ constraints. After setting mass of each leptoquark to $1.5\e{TeV}$ we find that of all considered states only $S_1$ leptoquark, coupled to both chiralities of leptons and quarks, is still a completely viable solution while the scenario with $R_2$ is in growing tension with $\Gamma(Z \to \tau \tau)$ and with the LHC constraints on the di-tau tails at high-$p_T$. We comment on the future experimental tests of $S_1$ scenario. 
\end{abstract}

\maketitle

\section{Introduction}

Experimental observations indicating the lepton flavor universality violation (LFUV) in the exclusive processes, based on transitions $b\to c\ell \nu$ and $b\to s\ell\ell$, provided a huge boost in the high energy physics community to build a model of physics beyond the Standard Model (BSM) which would capture the effects of LFUV while remaining consistent with many experimental tests of the Standard Model (SM). In that respect the scenarios in which the SM is extended by one or more $\mathcal{O}(1~\mathrm{TeV})$ leptoquarks proved to be the most attractive ones and a huge amount of work has been invested in (i) understanding the possible experimental signatures of such scenarios, and (ii) to figure out the ultraviolet~(UV) completion of the proposed models. A practical advantage of the models which involved only scalar leptoquarks is that the resulting theory remains renormalizable and therefore the loop processes can be computed and results compared with experiments. However, no single scalar leptoquark could describe both types of LFUV and therefore the viable models necessarily involved at least two scalar leptoquarks~\cite{Becirevic:2018afm,Crivellin:2019dwb,Gherardi:2020qhc,Becirevic:2022tsj,Iguro:2024hyk}.~\footnote{This last set of references also summarizes the evolution of model building with scalar leptoquarks in order to accommodate the so-called $R_{K^{(\ast)}}$ and $R_{D^{(\ast)}}$ anomalies.} An alternative to that situation was to use the simplest vector leptoquark, $U_1$, which could accommodate both types of LFUV, but in contrast to the models with scalar leptoquarks this scenario is not renormalizable and therefore the loop processes were explicitly dependent on the cutoff, which then necessitates to specify the UV completion, meaning new assumptions, new parameters (couplings and masses of additional states)~\cite{Buttazzo:2017ixm,Bordone:2017bld}. 
 
Recently, however, the hints of LFUV in the $b\to s\ell\ell$ modes, which for almost a decade indicated that the measured $R_{K^{(\ast )}} = \mathcal{B}(B\to K^{(\ast )} \mu\mu)/\mathcal{B}(B\to K^{(\ast )} ee) < 1$, were reexamined and the newly revised values established by LHCb were found to be, $R_K = 0.949(47)$ and $R_{K^{\ast}} = 1.027(75)$, thus fully consistent with lepton universality~\cite{LHCb:2022qnv}. 
The world average values of $R_{D^{(\ast )}} = \mathcal{B}(B\to D^{(\ast )} \tau\nu)/\mathcal{B}(B\to D^{(\ast )} l \nu)$, with $l\in \{e,\mu\}$, remain 
$R_{D^{(\ast )}}^\mathrm{exp}> R_{D^{(\ast )}}^\mathrm{SM}$ and a model with one $\mathcal{O}(1~\mathrm{TeV})$ scalar leptoquark is still a viable option to accommodate that experimental deviation. In this paper we revisit such models in a minimalistic setup. One, however,  has to acknowledge the emergence of another observable that showed a departure from its value predicted in the SM, namely $\mathcal{B}(B^\pm\to K^\pm \nu\bar\nu) =  2.40(67)\times 10^{-5}$~\cite{Belle-II:2023esi,Becirevic:2023aov}. Even though we do not intend to accommodate that deviation with respect to the SM prediction, we will monitor that our model does not lead to $\mathcal{B}(B^\pm\to K^\pm \nu\bar\nu)$ at odds with the experimental observations.

\section{Lepton universality violating effects in $b\to c \tau \nu$}

Before discussing the scalar leptoquark scenarios, we first consider the low-energy effective theory (LEET) of $b\to c \tau \nu$ transitions. We extend the usual LEET Lagrangian by including a singlet fermion $N_R$ (right-handed neutrino) because it is needed in the models involving the $\widetilde R_2$ leptoquark which can couple to $N_R$ at tree level. By focusing on the terms relevant to our paper we have:
\begin{align}
    \label{eq:LeffSL}
\mc{L}_\mrm{b\to c\tau \nu} = 
-2 \sqrt{2}  G_F V_{cb}&\Big[\left(1+g_{V_L}\right)\left(\bar{c}_{L} \gamma^\mu b_{L}\right)\left(\bar{\tau}_{L} \gamma_\mu \nu_{\tau L}\right)+g_{V_R}\left(\bar{c}_{R} \gamma^\mu b_{R}\right)\left(\bar{\tau}_{L} \gamma_\mu \nu_{\tau L}\right) \nonumber\\
& \quad 
+g_{S_L} \left(\bar{c}_{R} b_{L}\right)\left(\bar{\tau}_{R} \nu_{\tau L}\right)+g_T\left(\bar{c}_{R} \sigma^{\mu \nu} b_{L}\right)\left(\bar{\tau}_{R} \sigma_{\mu \nu} \nu_{\tau L}\right) + \\
&\quad +\widetilde g_{S_R} (\bar c_{L} b_{R})(\bar\tau_L N_R)+ \widetilde g_{T} (\bar c_{L}\sigma^{\mu\nu} b_{R})(\bar\tau_L \sigma_{\mu\nu}N_R)\Big] +\mathrm{h.c.}\,.\nonumber
\end{align}
In the framework of the SM effective theory (SMEFT), the $\tau$-specific $g_{V_R}$ can only stem from dimension-8 operator in the linear realisation of $SU(2)_L$ for the Higgs~\cite{Cata:2015lta} and we will not pursue this option in the rest of the paper.\footnote{E.g. $g_{V_R}$ can arise in a $R_2$-$\widetilde R_2$ model via mixing term between the two leptoquarks.}

\subsection{Matching to SMEFT}
\label{sec:SMEFT}

In the following we specify the matching and the renormalization group running between the SMEFT Wilson coefficients and the low-energy coefficients $g_i$ of Eq.~\eqref{eq:LeffSL}, defined at scale $\mu=m_b$. Above the electroweak scale the SMEFT Lagrangian consists of operators invariant under the full SM gauge group~\cite{Buchmuller:1985jz,Grzadkowski:2010es,Jenkins:2013wua,Alonso:2013hga}:
\begin{equation}
\label{eq:SMEFT-SL} 
    \mathcal{L}_\mathrm{SMEFT} = \frac{1}{\Lambda^2} \sum_i \, C_i \,\mc{O}_i\,.
\end{equation}
Since we consider models with a single leptoquark a natural choice for  normalization scale is $\Lambda=m_\mrm{LQ}$. For the semileptonic processes with the SM neutrino the following operators are relevant:
\begin{align}
    \label{eq:SMEFT-SLops}
    \mc{O}^{(1)}_{\substack{lequ\\prst}} &= (\bar l^a_p e_r) \epsilon^{ab} (\bar q^b_s  u_t)\,, \qquad \mc{O}^{(3)}_{\substack{lequ\\ prst}} = (\bar l^a_p \sigma^{\mu\nu} e_r) \epsilon^{ab} (\bar q^b_s  \sigma_{\mu\nu} u_t)\,,\\
    \mc{O}^{(1)}_{\substack{lq\\prst}} &= (\bar l^j_p \gamma^\mu l_r) (\bar q_s \gamma_\mu  u_t)\,, \qquad \mc{O}^{(3)}_{\substack{lq\\ prst}} = (\bar l_p \gamma^\mu \tau^I l_r)  (\bar q_s  \gamma_{\mu} \tau^I q_t)\,.
\end{align}
Here $l$ and $q$ denote the doublet lepton and quark fields, respectively, $u$ and $e$ are singlet up-quark and charged-lepton fields. Matrices $\tau^I$ are the Pauli-matrices ($I=1,2,3$) acting on $SU(2)_L$, while $a,b=1,2$ are the indices of $SU(2)_L$ doublets. Finally, the flavor indices are denoted by $prst$ and we employ the diagonal basis for the left handed down-type quarks and for charged leptons.
The presence of light $N_R$ requires to consider a more general effective theory, often referred to as $N_R$-SMEFT, that besides a full set of SMEFT operators includes all possible operators containing a singlet $N_R$~\cite{delAguila:2008ir,Fernandez-Martinez:2023phj}. As far as the contribution to $\rddst$ goes, we will employ two such $N_R$-SMEFT operators:
\begin{equation}
\label{eq:NRSMEFT}
\mc{O}^{(1)}_{\substack{Nldq \\ \phantom{N}rst}} = (\bar N_R l^a_r) \epsilon^{ab}(\bar d_s q_t^b)\,,\qquad
\mc{O}^{(3)}_{\substack{Nldq \\ \phantom{N} rst}} = (\bar N_R \sigma^{\mu\nu}l^a_r) \epsilon^{ab}(\bar d_s  \sigma_{\mu\nu}q_t^b)  \,.
\end{equation}
The literature on $N_R$-SMEFT considers operator $\mc{O}^{(1)}_{Nldq}$ and its version with exchanged roles of $l$ and $q$, $\mc{O}^{(1)}_{Nqdl}$. These two operators are related to those given in Eq.~\eqref{eq:NRSMEFT} via the Fierz identities.

The relations between SMEFT Wilson coefficients and the low-energy Wilson coefficients of Lagrangian~\eqref{eq:LeffSL} are obtained by the renormalization group running and matching of the two theories.
Vector current Wilson coefficients $C^{(1),(3)}_{lq}$ do not run and the coefficient $g_{V_L}$ is obtained as:
\begin{equation}
\label{eq:Vmatch}
g_{V_L}(m_b) = -\frac{v^2}{m_\mrm{LQ}^2} \frac{V_{cs}}{V_{cb}}C^{(3)}_{\substack{lq \\ \tau\tau s b}} (m_\mrm{LQ}) - \frac{v^2}{m_\mrm{LQ}^2} C^{(3)}_{\substack{lq \\ \tau\tau b b}} (m_\mrm{LQ})\,,
\end{equation}
where we have assumed both possibilities of quark flavors at high energy scales.
On the other hand, scalar and tensor operators $C_{lequ}^{(1)}$ and $C_{lequ}^{(3)}$ renormalize and mix among themselves.
The low energy Wilson coefficients present in~\eqref{eq:LeffSL} are obtained by the renormalization group running and matching onto the low energy effective theory~\eqref{eq:LeffSL}
 \begin{align}
\label{eq:STmatchandrun}
\begin{split}
   g_{S} (m_b)  &=   -\frac{v^2}{2 m_\mrm{LQ}^2} \frac{1}{V_{cb}} \left(\frac{\alpha_s(m_b)}{\alpha_s(m_t)}\right)^{12/23} \left(\frac{\alpha_s(m_t)}{\alpha_s(m_\mathrm{LQ})}\right)^{4/7}  C^{(1)*}_{\substack{lequ\\ \tau\tau b c}}(m_\mrm{LQ})
        =-0.56\, C^{(1)*}_{\substack{lequ\\ \tau\tau b c}}(m_\mrm{LQ}) \,,\\
        g_{T} (m_b)&=  -\frac{v^2}{2 m_\mrm{LQ}^2} \frac{1}{V_{cb}} \left(\frac{\alpha_s(m_b)}{\alpha_s(m_t)}\right)^{-4/23} \left(\frac{\alpha_s(m_t)}{\alpha_s(m_\mathrm{LQ})}\right)^{-4/21}   C^{(3)*}_{\substack{lequ\\ \tau\tau b c}}(m_\mrm{LQ})= -0.28\,C^{(3)*}_{\substack{lequ\\ \tau\tau b c}}(m_\mrm{LQ})\,,
\end{split}
\end{align}
obtained to leading order in QCD. The off-diagonal mixings, i.e. mixing of $C^{(3)}_{lequ}$ to $g_{S_L}$ and $C^{(1)}_{lequ}$ to $g_{T}$, are driven by electroweak radiative contributions and represent only small corrections with respect to the QCD driven multiplicative renormalization. 
The invariance of QCD under parity implies that the chirality-flipped operators involving right-handed $N_R$ have exactly the same QCD running, i.e.:
\begin{align}
\label{eq:STtildematchandrun}
\begin{split}
    \widetilde g_{S_R} (m_b)  &=
    -0.56\, C^{(1)*}_{\substack{Nldq \\ \tau bs}}(m_\mrm{LQ}) \,, \\ 
        \widetilde g_{T} (m_b)&=   
      -0.28\,  C^{(3)*}_{\substack{Nldq\\ \tau b s}}(m_\mrm{LQ}) \,.
\end{split}
\end{align}
The above discussion is based on 1-loop running only, but in actual computations we included the higher corrections to 4-loops~\cite{Chetyrkin:1997dh,Vermaseren:1997fq,Gracey:2022vqr,vanRitbergen:1997va,Chetyrkin:1997sg}. The net effect is that the right hand side of Eq.(\ref{eq:STmatchandrun}) for  $g_{S_L}$ ($g_{T}$) gets enhanced (suppressed) by about $6\%$ ($4\%$). In other words, the relation $g_{S_L} = \pm 4\, g_{T}$ valid at the $\mu=m_\mrm{LQ}$,  becomes  $g_{S_L} = \pm 8.8\, g_{T}$ when the 4-loop running is used.\footnote{Notice that the effect of 1-loop running at $\mu=m_b$ is $g_{S_L} = \pm 8\, g_{T}$ . }
The same applies to Eq.~\eqref{eq:STtildematchandrun} regarding the couplings $\widetilde g_{S_R}$ and $\widetilde g_{T}$.

\subsection{$R_D$ and $R_{D^*}$}
The goal of this study is to establish whether or not any of the scalar leptoquarks, with a minimalistic set of Yukawa couplings, fits the current experimental world average of $R_{D}$ and $R_{D^*}$. 
The experimental averages are steadily updated in Ref.~\cite{HFLAV:2022esi}. The most recent values are:
\begin{equation}\label{eq:Rexp}
R_{D}^\mathrm{exp} = 0.344(26)\,,\qquad R_{D^*}^\mathrm{exp} = 0.285(12)\,,
\end{equation}
with a correlation coefficient $\rho=-0.39$.

The SM predictions for these quantities are plagued by systematic uncertainties arising from hadronic matrix elements, i.e. from the relevant form factors. In the case of $D$-meson in the final state the problem is easier to handle because only the vector current contributes, i.e. only two form factors are present, $f_{+,0}(q^2)$, namely,
\begin{align}
\label{eq:ff}
\langle D(k) | \bar{s}\gamma^\mu b | B(p) \rangle &= \Big{[}(p+k)^\mu- \dfrac{m_B^2-m_D^2}{q^2}q^\mu\Big{]} f_{+}(q^2) +\dfrac{m_B^2-m_D^2}{q^2} q^\mu f_0(q^2)\,,
\end{align}
which are equal at $q^2 = 0$, $f_+(0)=f_0(0)$, where $q=p-k$ and $q^2 \in [m_\ell^2, (m_B-m_D)^2]$, with $\ell \in \{e,\mu,\tau\}$. That condition is extremely helpful when extrapolating the form factors that are accessible through lattice QCD at large $q^2$'s down to low $q^2$'s. Two such lattice QCD analyses~\cite{MILC:2015uhg,Na:2015kha} agree and their results are combined in Ref.~\cite{FlavourLatticeAveragingGroupFLAG:2021npn}, which is used to make the SM prediction,
\begin{equation}\label{eq:RDSM}
R_D^\mathrm{SM}= 0.293(8)\,,
\end{equation}
which is a little less than $2\sigma$ smaller than measured, cf. Eq.~\eqref{eq:Rexp}.
For the New Physics contributions to the $B \to D \tau \nu$ amplitude we will also employ the tensor form factor $f_T(q^2)$, defined as:
\begin{equation}
    \Braket{D(k) | \bar{s}\sigma^{\mu \nu} b | B(p) } = \frac{-2 i\, f_T(q^2)}{m_B+m_D}  (p^\mu k^\nu - p^\nu k^\mu)\,.
\end{equation}
Note that $f_T$ also depends on the renormalization scale $\mu$. It was observed that shapes of $f_T(q^2)$ and $f_+(q^2)$ are similar, compatible with $q^2$-independent ratio of the two form factors $f_T(q^2)/f_+(q^2) = 1.06(12)$~\cite{Atoui:2013zza}.

The situation with $R_{D^*}$ is more complicated. Both the vector and the axial currents contribute to the $B\to D^\ast \ell\bar \nu$ decay amplitude, which then involves $4$ independent form factors:
\begin{equation}
\begin{split}\label{eq:ffDstar}
\langle D^\ast(k) | \bar{c}\gamma_\mu(1-\gamma_5) b &| B(p) \rangle = \varepsilon_{\mu\nu\rho\sigma} \varepsilon^{\ast\nu}p^\rho k^\sigma \dfrac{2 V(q^2)}{m_B+m_{D^\ast}} -i\varepsilon_\mu^\ast(m_B+m_{D^\ast})A_1(q^2) \\
+ &i(p+k)_\mu (\varepsilon^\ast\cdot q)\dfrac{A_2(q^2)}{m_B+m_{D^\ast}}
+i q_\mu (\varepsilon^\ast\cdot q) \dfrac{2 m_{D^\ast}}{q^2} \left[A_3(q^2)-A_0(q^2)\right]\,,
\end{split}
\end{equation}
where $\varepsilon_\mu$ is the $D^*$ polarization vector. Notice that the form factor $A_3(q^2)$ is not independent but related to $A_{1,2}(q^2)$, via 
$2 m_{D^\ast} A_3(q^2) = (m_B+m_{D^\ast}) A_1(q^2)-(m_B- m_{D^\ast}) A_2(q^2)$. It is also related to $A_0(q^2)$ at $q^2=0$ as $A_3(0)=A_0(0)$. 
 We prefer the parametrization given in Eq.~\eqref{eq:ffDstar} to the alternative one occasionally used in literature~\cite{Boyd:1997kz}, because the form factors are dimensionless by definition~\eqref{eq:ffDstar}. Physical results are obviously independent of the parametrization used. 
The fact that there are many more form factors and only one constraint [$A_3(0)=A_0(0)$] make the lattice QCD computation of the $B\to D^\ast$ form factors particularly challenging when extrapolating the results obtained for a few small three-momenta given to the lighter meson down to lower end of $q^2 = (p-k)^2 \in [m_\ell^2, (m_B-m_{D^\ast})^2]$. Very recently, the results of three detailed lattice QCD computations of these form factors have been reported in Refs.~\cite{FermilabLattice:2021cdg,Harrison:2023dzh,Aoki:2023qpa}. When converted to the same set of form factors, such as those defined in Eq.~\eqref{eq:ffDstar}, it appears that they are quite consistent when it comes to the dominant form factor, $A_1(q^2)$, but they do not agree for the other form factors. That disagreement, when extrapolated to low $q^2$'s can lead to very different predictions. To avoid such a situation, we will proceed as follows. Since both quarks participating in this process are heavy, $m_{c,b} \gg \Lambda_\mathrm{QCD}$, it is reasonable to use the decomposition of  hadronic matrix elements in terms of the form factors motivated by the heavy quark effective theory, namely, 
\begin{equation}
\begin{split}\label{eq:ffHQE}
\frac{1}{\sqrt{m_B m_{D^\ast}}}\Braket{D^\ast(k) | \bar{c}\gamma_\mu(1-\gamma_5) b| B(p) } &=  \varepsilon_{\mu\nu\rho\sigma} \varepsilon^{\ast\nu}v^\rho v^{\prime \sigma} h_V(w) -i\varepsilon_\mu^\ast(w+1)h_{A_1}(w) \\
&\quad+ i (\varepsilon^\ast\cdot v) v_\mu h_{A_2}(w) 
+i (\varepsilon^\ast\cdot v) v^\prime_\mu  h_{A_3}(w) \,, 
\end{split}
\end{equation}
which are used to fit the experimentally measured angular distributions of the $B\to D^\ast (\to D\pi) l\bar \nu$ decay ($l\in\{e,\mu\}$). In that way the normalization and shapes of the above form factors is reconstructed  from the measured angular distribution. In the above expressions, $v_\mu = p_\mu/m_B$, $v^\prime_\mu = k_\mu/m_{D^\ast}$, and $w=v\cdot v^\prime = (m_B^2+m_{D^\ast}^2 - q^2)/(2 m_B m_{D^\ast})$. More specifically, with the educated assumptions regarding the shapes, as proposed in Ref.~\cite{Caprini:1997mu}, the following expressions have been used:
\begin{align}\label{eq:CLN}
\begin{split}
h_{A_1}(w)&=  h_{A_1}(1) \, \left[
1 - 8 \rho^2 z + (53 \rho^2 -15) z^2 - (231 \rho^2 - 91) z^3 \right]
\,, \\
R_1(w)& =\frac{h_V(w)}{h_{A_1}(w)} = R_1(1) - 0.12 (w-1) + 0.05 (w-1)^2\,, \\
R_2(w)& =\frac{ h_{A_3}(w) + (m_{D^\ast}/m_B) h_{A_2}(w)}{h_{A_1}(w)}  = R_2(1) + 0.11 (w-1) - 0.06 (w-1)^2\,,
\end{split}
\end{align}
where $z= (\sqrt{w+1}-\sqrt{2})/(\sqrt{w+1}+\sqrt{2})$.
Therefore, only four parameters are to be obtained from the fit with experimental data: $h_{A_1}(1)$, $\rho^2$, $R_{1}(1)$ and $R_{2}(1)$. After combining the experimental results of several experiments, cf. Ref.~\cite{HFLAV:2022esi}, the following values have been quoted: 
\begin{equation}
\label{eq:paramsDstar}
\rho^2 = 1.121(24), \quad R_1(1)=1.269(26),\quad R_2(1)=0.853(17),
\end{equation}
while the value of $h_{A_1}(1)$, being an overall multiplicative factor, is immaterial for the discussion of $R_{D^\ast}$. The correlation matrix for these parameters is given in Ref.~\cite{HFLAV:2022esi} and we used it in our numerical estimates. Note that the Belle~II results of the above parameters~\cite{Belle:2023bwv} appeared after the release of the HFLAV review~\cite{HFLAV:2022esi} but the reported results are fully consistent with the numbers quoted in Eq.~\eqref{eq:paramsDstar}. 

The relations between the form factors given in Eqs.~(\ref{eq:ffHQE},\ref{eq:CLN}) and those in~\eqref{eq:ffDstar} are: 
\begin{align}\label{eq:FFrelations}
A_1(q^2) & = \frac{\sqrt{m_B m_{D^*}}}{m_B + m_{D^*}} \left( w +1\right)\, h_{A_1}\left( w\right)\,, \nonumber\\
\frac{V(q^2)}{A_1(q^2)} & =\left[ 1 - \frac{q^2}{(m_B + m_{D^*})^2}\right]^{-1} R_1\left( w\right)\,,\\
\frac{A_2(q^2)}{A_1(q^2)} & =\left[1 - \frac{q^2}{(m_B + m_{D^*})^2}\right]^{-1} R_2\left( w\right)\,,\nonumber
\end{align}
where, again, $w\geq 1$ is the recoil momentum, $q^2=m_B^2+m_{D^*}^2 - 2 m_B m_{D^*}\,w$. The values of parameters given in Eq.~\eqref{eq:paramsDstar} are obtained from experimental studies of the angular distribution of $B\to D^\ast (\to D\pi) l\bar \nu$, with $l\in\{e,\mu\}$, for which $m_l^2/m_{B,D}^2$ is negligibly small and therefore the pseudoscalar form factor $A_0(q^2)$ cannot be accessed. That form factor, however, contributes significantly to 
$B\to D^\ast \tau\bar \nu$. To get that information we will define
\begin{equation}
R_0(w) = \left[ 1 - \frac{q^2}{(m_B + m_{D^*})^2}\right] \frac{A_0(q^2)}{A_1(q^2)}\,,
\end{equation}
which is known at $w_\mathrm{max}$ (i.e. $q^2=0$), due to 
\begin{equation}
\frac{A_0(0)}{A_1(0)}  =
\frac{1}{2m_{D^*}}\left[ m_B+ m_{D^*} - (m_B- m_{D^*}) \frac{A_2(0)}{A_1(0)} \right]\,.
\end{equation}
We thus have 
\begin{equation}\label{eq:wmax}
R_0(w_\mathrm{max}) = \frac{m_B + m_{D^*}}{2 m_{D^*}} - \frac{m_B - m_{D^*}}{2 m_{D^*}} R_2(w_\mathrm{max}) =1.087(14),
\end{equation}
where we used information from Eqs~(\ref{eq:CLN},\ref{eq:paramsDstar}) to get the
last number. We need at least one extra point to have the information on the slope of the form factor ratio  $R_0(w)$. To that end we can use the lattice QCD results which are computed at small $w$ values (or equivalently, large $q^2$'s). From the information provided in their papers, at $w=1$, we find:  $A_0/A_1 = 1.423(53), 1.387(49), 1.186(72)$ for MILC/FNAL~\cite{FermilabLattice:2021cdg}, JLQCD~\cite{Aoki:2023qpa} and HPQCD~\cite{Harrison:2023dzh}, respectively. After taking the average, 
\begin{equation}
R_0(1) = \frac{4 m_B  m_{D^*}}{(m_B + m_{D^*})^2} \frac{A_0(q^2_\mathrm{max})}{A_1(q^2_\mathrm{max})} =1.087(26)\,,
\end{equation}
which, when compared with Eq.~\eqref{eq:wmax}, indicates a flat behavior of $R_0(w)$. If, instead, we took separately the average of lattice values for $A_0(q^2_\mathrm{max})$ and of $A_1(q^2_\mathrm{max})$, and then combined them in the ratio, we would have obtained $R_0(1) = 1.095(31)$,
indicating a small negative slope in $w-1$. To capture these effects we will take:
\begin{equation}\label{eq:R0number}
R_0(w) = 1.09 - 0.16 (w-1)\,
\end{equation}
to which we attribute an error of $10\%$, to account for the spread of lattice data at $w=1$. With the ingredients described above we obtain
\begin{equation}\label{eq:RDstarSM}
R_{D^*}^\mathrm{SM} = 0.247(2)\,,
\end{equation}
hence over $3\sigma$ smaller than the current experimental average given in Eq.~\eqref{eq:Rexp}.

We should stress once again that our basic assumption is that the BSM physics can modify the coupling to $\tau$-lepton while leaving the couplings to lighter leptons intact. This is why we could use the experimental information on the form factors. If that assumption is relaxed then one ends up with very large error bar on $R_{D^\ast}$, reflecting the disagreement among various lattice QCD results~\cite{Martinelli:2023fwm,Martinelli:2021onb}. To consider the BSM scenarios captured by the Lagrangian~\eqref{eq:LeffSL} we need three more form factors $T_{1,2,3}(q^2)$ defined as, 
\begin{equation}
\label{eq:tensorFF}
\begin{split}
\Braket{ D^\ast(k)|\bar c \sigma_{\mu\nu} b|B(p) } &= \,\epsilon_{\mu\nu\alpha\beta} \left\{ -\varepsilon^{*\alpha}(p+k)^\beta T_1(q^2) + \varepsilon^{*\alpha}q^\beta 
\frac{m_B^2-m_{D^\ast}^2}{q^2}[T_1(q^2)-T_2(q^2)] 
\right. \\
& \quad\left. \qquad + (\varepsilon^* q)p^\alpha k^\beta \frac{2}{q^2} \left[ T_1(q^2)-T_2(q^2) - \frac{q^2}{m_B^2 - m_{D^\ast}^2} T_3(q^2) \right] \right\} \,. 
\end{split}
\end{equation}
They have been very recently computed in Ref.~\cite{Harrison:2023dzh}. From that paper we extract (at $\mu=m_b$):
\begin{align}\label{eq:FFrelations2}
\frac{T_1(q^2)}{A_1(q^2)} & = 0.96 + 0.041 \, q^2/\mrm{GeV}^2\,,\nonumber \\
\frac{T_2(q^2)}{A_1(q^2)} & = 0.96 - 0.025 \, q^2/\mrm{GeV}^2 - 0.002 \, q^4/\mrm{GeV}^4\,,\\
\frac{T_3(q^2)}{A_1(q^2)} & = 0.35 + 0.040 \, q^2/\mrm{GeV}^2  \,.\nonumber
\end{align}
with an overall error of $6\%$, $4\%$ and $30\%$, respectively.

With the information given above it is convenient to write:
\begin{align}
\label{eq:magic}
\begin{split}
\frac{R_{D^{(\ast )}}}{R_{D^{(\ast )}}^{\mathrm{SM}}} \, =\, &\left|1+g_{V_L}\right|^2 +a_S^{D^{(\ast )}}\left(\left|g_{S_L}\right|^2 +\left|\widetilde g_{S_R}\right|^2\right)
+a_T^{D^{(\ast )}} \left(\left|g_T\right|^2 + \left|\widetilde g_T\right|^2\right)\\
&\quad + a_{S V}^{D^{(\ast )}} \, \re\left[(1+g_{V_L})\, g_{S_L}^*\right] + a_{T V}^{D^{(*)}} \, \re \left[ (1+g_{V_L}) \, g_T^*\right]   \,, 
\end{split}
\end{align}
where in the case of $D$ in the final state, thanks to the lattice QCD results~\cite{Na:2015kha,MILC:2015uhg,Atoui:2013zza}, we obtain~\cite{Feruglio:2018fxo,Iguro:2022yzr}:
\begin{align}
    a_S^D= 1.08(1),\quad 
    a_T^D=0.83(5),\quad 
    a_{SV}^D=1.54(2),\quad
    a_{TV}^D=1.09(3).
\end{align}

Instead, for the case of $D^*$ in the final state, and based on the discussion presented above, we have:
\begin{align}\label{magicDstar}
    a_S^{D^*}= 0.037(4),\quad 
    a_T^{D^*}=8.56(35),\quad 
    a_{SV}^{D^*}=-0.107(11),\quad
    a_{TV}^{D^*}=-2.91(11)\,.
\end{align}
Note that the lattice QCD results of the form factors $T_{1,2,3}$  significantly reduced the values of the coefficients $a_T^{D^*}$ and $a_{TV}^{D^*}$, when compared to the values given in Refs.~\cite{Feruglio:2018fxo,Mandal:2020htr,Iguro:2022yzr}.

Before closing this Section we should reiterate that our estimates of $R_{D^\ast}^\mathrm{SM}$~\eqref{eq:RDstarSM} and of the coefficients in  Eq.~\eqref{magicDstar} are obtained by mostly relying on the form factors obtained from the experimental data which were averaged and fit to the parametrization of Ref.~\cite{Caprini:1997mu}, also given in Eq.~\eqref{eq:CLN}. Of course that parametrization is not the only one possible and it already received criticism in the literature, suggesting that its extension proposed in Ref.~\cite{Bernlochner:2017jka} or an alternative parametrization of Ref.~\cite{Boyd:1997kz} should be used. With the current precision of experimental data, however, both parametrizations fit equally well the data and the corresponding results are fully compatible. To make that point more explicit we used the results presented in the most recent experimental analysis of the angular distribution of $B\to D^\ast (\to D\pi ) l \nu$, presented in Ref.~\cite{Belle:2023bwv}, in which the form factors are extracted by using both parametrizations, namely the CLN of Ref.~\cite{Caprini:1997mu} and the BGL one of Ref.~\cite{Boyd:1997kz}. Both parametrizations give a satisfactory $\chi^2/\mathrm{d.o.f.}$ and therefore lead to the same value of $R_{D^\ast}^\mathrm{SM}$, as well as to almost indistinguishable values of the parameters  $a_{S,T,SV,TV}^{D^*}$.~\footnote{To be fully explicit, we find $R_{D^\ast}^\mathrm{SM}=0.255(5)$ and 
\begin{align}
 & \underline{\mathrm{CLN-Belle~II}}: 
 \quad  a_S^{D^*}= 0.039(7),\quad 
    a_T^{D^*}=8.41(64),\quad 
    a_{SV}^{D^*}=-0.111(20),\quad
    a_{TV}^{D^*}=-2.80(20)\,,\nonumber\\
&   \underline{\mathrm{BGL-Belle~II}}:
\quad   a_S^{D^*}= 0.039(7),\quad 
    a_T^{D^*}=8.46(64),\quad 
    a_{SV}^{D^*}=-0.110(21),\quad
    a_{TV}^{D^*}=- 2.84(19)\,, \nonumber
\end{align}
where in the evaluation of the above coefficients, besides the form factors extracted from the Belle~II data~\cite{Belle:2023bwv}, we used Eqs.~(\ref{eq:R0number},\ref{eq:FFrelations2}).}

\subsection{Collider constraints on leptoquarks}

As already mentioned above, we attribute the deviations of the observed $R_{D^{( \ast)}}$~\eqref{eq:Rexp} with respect to their values predicted in the SM~(\ref{eq:RDSM},\ref{eq:RDstarSM}) to a coupling of new physics particles to the third generation of leptons. In particular, we focus on the scenarios in which the scalar leptoquark couples to $\tau$ and either to $c$- or to $b$-quark. We refer to those coupling as to Yukawa couplings. One should monitor that such couplings do not become too large so that they could result in a significant modification of the high dilepton mass tails of $pp \to \tau \nu, \tau \tau$ processes~\cite{Eboli:1987vb,Faroughy:2016osc,Alves:2018krf,Greljo:2018tzh,Iguro:2020keo}. Both ATLAS~\cite{2002.12223,ATLAS:2021bjk} and CMS~\cite{2208.02717} have presented results of their studies of such Drell-Yan processes at high dilepton masses. In this work we use the HighPT package~\cite{Allwicher:2022mcg,2207.10714} which provides us with a built-in likelihood function of the leptoquark couplings for each of the leptoquarks (for high-$p_T$ constraints in the case of $S_1$ leptoquark cf.~\cite{Mandal:2018kau}). 
They are obtained after recasting the results of Refs.~\cite{2002.12223,ATLAS:2021bjk} to the scenarios discussed in this paper. The exception is $\widetilde R_2$ and its coupling to $N_R$ which has not been considered in HighPT. In Sec.~\ref{sec:R2t} we will explain  how one can simulate Drell-Yan effects of $\widetilde R_2$ by those of $R_2$ coupled to ordinary neutrinos.

If light enough the leptoquarks can be produced in pairs via QCD interactions. Single leptoquark production, on the other hand, 
is completely determined by the Yukawa couplings. To satisfy the upper bounds on the leptoquark-mediated cross sections determined by ATLAS and CMS we set the leptoquark mass to $m_\mrm{LQ} = 1.5\e{TeV}$ for all cases~\cite{ATLAS:2021oiz,ATLAS:2024huc,CMS:2023qdw}.

\subsection{$B \to K^{(*)} \nu \bar \nu$}
While our goal is not to accommodate the recently observed deviation of the measured $\mathcal{B}(B \to K \nu \bar \nu)$~\cite{Belle-II:2023esi} with respect to the SM prediction~\cite{Becirevic:2023aov,Allwicher:2023xba}, we should monitor that our scenarios do not get in conflict with the experimental bounds on $\mathcal{B}(B \to K^{(\ast)} \nu \bar \nu)$~\cite{ParticleDataGroup:2022pth}. 
To do so we consider the following effective Lagrangian:\begin{align}
  \label{eq:bsnunu}
\mathcal{L}_{\mathrm{eff}}^{b\to s\nu\nu}= \dfrac{\sqrt{2}G_F \alpha_{\mathrm{em}} \lambda_t}{\pi}&\left[ C_L^{ij}~\big{(}\bar{s}_L\gamma_\mu b_L\big{)}\big{(}\bar{\nu}_{L\,i} \gamma^\mu \nu_{L\,j} \big{)} + C_R^{ij}~\big{(}\bar{s}_R\gamma_\mu b_R\big{)}\big{(}\bar{\nu}_{L\,i} \gamma^\mu \nu_{L\,j} \big{)} \right.\,,\nonumber\\
&\left. + \widetilde C_L^{NN}~\big{(}\bar{s}_L\gamma_\mu b_L\big{)}\big{(}\bar{N}_{R} \gamma^\mu N_{R} \big{)} + \widetilde C_{LR}^{iN}~\big{(}\bar{s}_L b_R\big{)}\big{(}\bar{\nu}_{Li} N_{R} \big{)} \right.\\
&\left. + \widetilde C_{RL}^{Ni}~\big{(}\bar{s}_R b_L\big{)}\big{(}\bar{N}_{R} \nu_{Li} \big{)} \right]+ \mrm{h.c.}\,,\nonumber
\end{align}
where $i=e,\mu,\tau$. The SM contribution is characterized by a unique and flavor diagonal left-handed interaction, $C_{L,\mrm{SM}}^{ij} = C_{L,\mrm{SM}} \delta^{ij}$, with $C_{L,\mrm{SM}} = -6.32(7)$. Leptoquarks, considered in this work can contribute to operators with different Lorentz structures or with $N_R$ included in the interactions.
We will use the expressions and ingredients from Refs.~\cite{Becirevic:2023aov,Allwicher:2023xba} to check on the experimental bounds~\cite{Belle-II:2023esi,ParticleDataGroup:2022pth}.

\subsection{Loop-induced constraints: $Z \to \ell \ell, \nu \nu$ and $\tau \to l \nu \bar \nu$}
Since accommodating $\rddst^\mathrm{exp}>\rddst^\mathrm{SM}$ requires significant coupling of the thrid family of leptons to leptoquarks, we should monitor that the resulting $\mathcal{B}(Z\to \tau \tau)$ and $\mathcal{B}(Z\to \nu \bar \nu)$ remain within the experimental error bars\cite{ParticleDataGroup:2022pth}. To do so we employ the
expressions derived in Ref.~\cite{Arnan:2019olv} for all of the scalar leptoquarks and confront them with measured effective couplings $g_{V,A}^{\tau}$ of $Z$ boson. Similarly, the couplings to neutrinos should remain consistent with $N_\mrm{eff} = 2.9840(82)$~\cite{ALEPH:2005ab}. 

Modifications of the on-shell fermionic couplings to $W$ are less precisely known and we do not consider them here. On the other hand, if the quark loop is charged (e.g. $\bar t b$ or $\bar c b$), it will induce, via $W$-exchange, leptonic decays of $\tau$~\cite{Feruglio:2016gvd,Arnan:2019olv}, which we do consider in our analyses.

\section{Leptoquark models}

In this Section we focus on three specific scenarios of SM extended by a presence of a single scalar $\mathcal{O}(1\,\mathrm{TeV})$ leptoquark that could provide us with a plausible explanation of $\rddst^\mathrm{exp}>\rddst^\mathrm{SM}$ through couplings to the third generation of leptons. In doing so we consider the models with a minimal number of Yukawa couplings. Three such scenarios allow for couplings to $b\tau$ and/or $c\tau$ and will be discussed one by one in the following.

\subsection{$R_2$}
In terms of the SM quantum numbers,\footnote{In the notation we employ the quantum numbers correspond to $(\mathrm{SU}(3),\mathrm{SU}(2)_L)_{\mathrm{U}(1)_Y}$ of the SM gauge group.} the $R_2$ doublet of scalar leptoquarks corresponds to $(3,2)_{7/6}$, so that the electric charge of its components is $Q=2/3$ and $Q=5/3$. 
The Yukawa interactions are described via
\begin{equation}
  \label{eq:R2Lagr}
  \mathcal{L}_{R_2} =  y_{R}^{ij}\, \bar{Q}_i^a e_j R_2^a\ +\  y_L^{ij}\, \bar{u}_{Ri} R_2^{T,a} \epsilon^{ab}L_j^b + \mrm{h.c.}\,,
\end{equation}
where $Q$ and $L$ are the quark and lepton doublets, i.e. $Q_i = [(V^{\dagger}u)_i,  d_i]^T$, $L = [\nu_\ell, \ell]^T$, with $V$ being the Cabibbo--Kobayashi-Maskawa (CKM) matrix. Notice that the left-handed neutrinos are in the flavor basis.
The key element in building a model is to specify the relevant Yukawa couplings. We opt for minimal number of parameters and in the down-quark and charged-lepton mass basis we choose: 
\begin{equation}
  \label{eq:YukR2}
  y_R =
  \begin{pmatrix}
    0 & 0 & 0\\
    0 & 0 & 0\\
    0 & 0 & y_R^{b\tau}
  \end{pmatrix}\,, \qquad y_L = \begin{pmatrix}
    0 & 0 & 0\\
    0 & 0 & y_L^{c\tau} \\
    0 & 0 & 0
  \end{pmatrix} \,,
\end{equation}
which leads to the interaction Lagrangian:
  \begin{align}
  \label{eq:LR2mass}
  \mc{L}_{R_2} &= y^{b\tau}_R V^*_{j b} (\overline{u}_j P_R \tau) R^{5/3}_2 + y^{b\tau}_R (\overline{b} P_R \tau) R^{2/3}_2 
  -y_L^{c\tau} (\overline{c} P_L \tau) R^{5/3}_2 + y_L^{c\tau} (\overline{c} P_L \nu_{\tau}) R_2^{2/3} + \mathrm{h.c.}\,.
  \end{align}
This $R_2$ model brings a tree level contribution to $\rddst$ which, when matched to SMEFT at the scale $\mu = m_{R_2} = 1.5~\e{TeV}$, is parametrized by two operators, $\mc{O}^{(1)}_{lequ}$ and $\mc{O}^{(3)}_{lequ}$, the coefficients of which satisfy
\begin{align}
\label{eq:R2treematching}
  C^{(1)}_{\substack{lequ\\ \tau\tau b c}}(m_{R_2}) & = 4\, C^{(3)}_{\substack{lequ\\ \tau\tau b c}}(m_{R_2}) = -\frac{y_R^{b\tau} {y_L^{c \tau}}^* }{2}\,. 
\end{align}
In terms of the LEET~\eqref{eq:LeffSL} the relevant couplings are $g_{S_L}$ and $g_T$. Their relation at the matching scale~\eqref{eq:R2treematching} $g_{S_L}(m_{R_2}) = 4 g_T(m_{R_2})$ is modified by the effects of running down to $\mu = m_b$~\eqref{eq:STmatchandrun}, and reads:
\begin{equation}\label{eq:gST}
    g_{S_L}(m_b)\ = \ 8.8\times g_T(m_b)\,.
\end{equation}

\begin{figure}
    \centering
    \includegraphics[scale=.78]{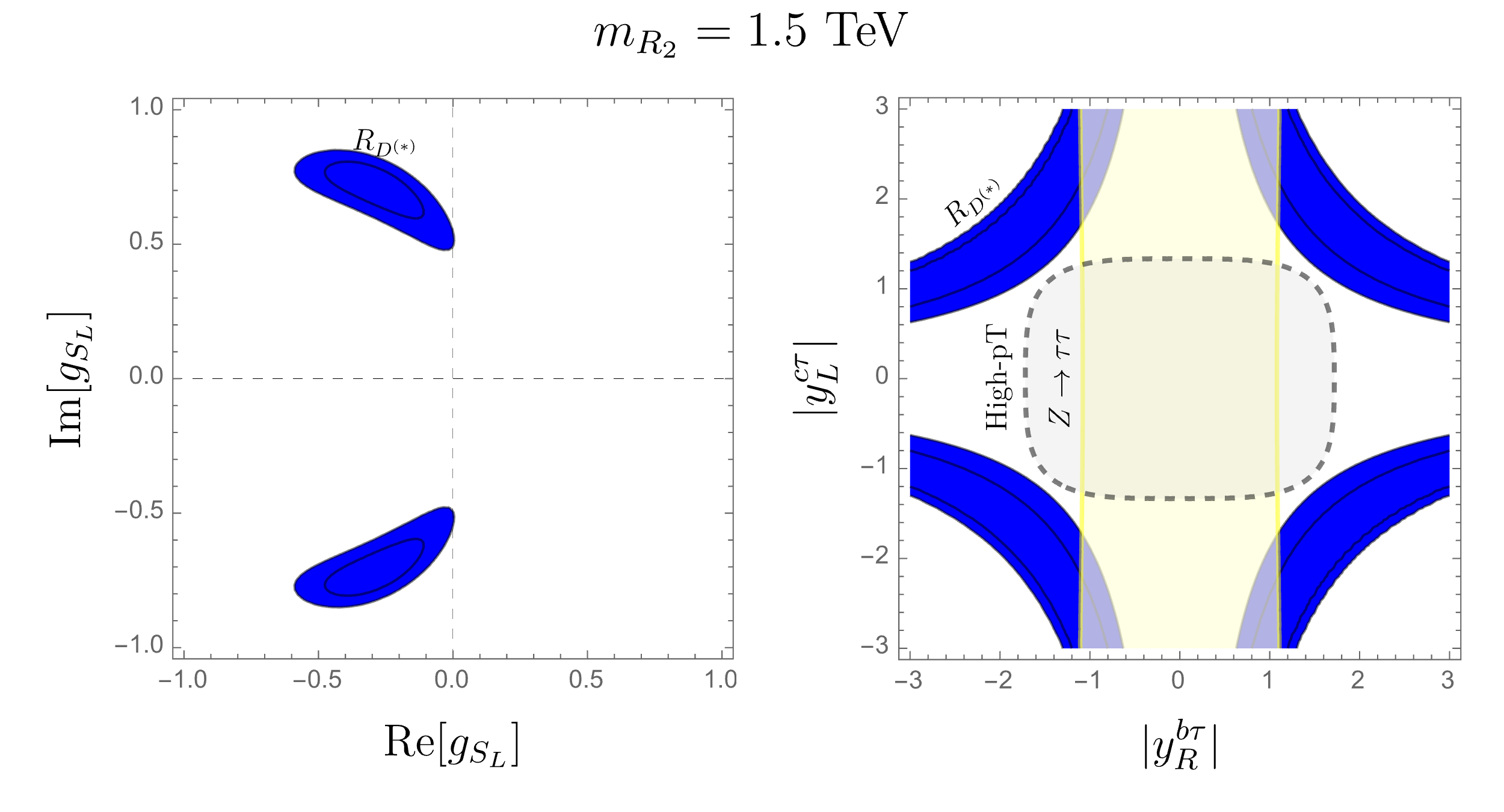}
    \caption{\small \sl $R_2$ bounds: In the left plot are shown the real and imaginary parts of $g_{S_L}(\mu=m_b)$ compatible with $R_{D^{(\ast)}}^\mathrm{exp}$ to 1 and $2\sigma$. The same color is then used to show that same  constraint on the Yukawa couplings in the right plot, but at the scale $\mu=m_{R_2}$ which we choose to be $m_{R_2}=1.5$ TeV. The vertical band in the right plot stems from the $2\sigma$ consistency with the measured $\mathcal{B}(Z\to \tau\tau)$. Inside the dashed is the region allowed by the experimental studies of high-$p_T$ tails of $pp\to \tau\nu,\tau\tau$, again to $2\sigma$. }
    \label{fig:bounds_R2} 
\end{figure}

It is well known that accommodating $\rddst^\mathrm{exp}>\rddst^\mathrm{SM}$ in a scenario with $R_2$ necessitates introducing a complex coupling $g_{S_L}(\mu=m_b)$~\cite{Sakaki:2013bfa,Becirevic:2018afm,Cheung:2020sbq,Becirevic:2022tsj}, in a way consistent with Eqs.~(\ref{eq:magic},\ref{eq:gST}). This is shown in the left panel of Fig.~\ref{fig:bounds_R2}. That also means that the product of two Yukawa couplings has to be complex,\footnote{We choose to attribute the complex phase to $y_R^{b\tau} = |y_R^{b\tau}| e^{i\varphi}$. For alternative constraints on this CP-violating phase see Ref.~\cite{Becirevic:2022tsj}.}
\begin{equation}
    g_{S_L}(m_b) = 0.60\times\frac{1}{2}\, |y_R^{b\tau} y_L^{c\tau}|e^{i\varphi}\,\,.
\end{equation}
The best fit values to $\rddst$ are
\begin{equation}
   |g_{S_L}(m_b)|= 0.78\,,\qquad   \varphi=\pm 1.96\,(=\pm 112^{\circ})\,.
\end{equation}

In the right panel of Fig.~\ref{fig:bounds_R2} we show the other important constraints, but this time in the plane span by the moduli of our Yukawa couplings ($|y_R^{b\tau}|$,$|y_L^{c\tau}|$). We see that the $2\sigma$ constraints arising from experimental studies of the di-tau and mono-tau high-$p_T$ tails at the LHC are at odds with the values of Yukawa couplings preferred by $R_{D^{(\ast )}}^\mathrm{exp}$. Otherwise the $2\sigma$ constraint stemming from consistency with the measured $\Gamma(Z\to \tau\tau)$ [or, better, $g_{V,A}^\tau$~\cite{ParticleDataGroup:2022pth}] would select larger values of  $|y_L^{c\tau}|$ while keeping moderately small $|y_R^{b\tau}|$. 
We reiterate that we use $m_{R_2}=1.5$~TeV, consistent with the lowest mass allowed for a leptoquark decaying mostly to $c\tau$ which is experimentally set to be $1.3$~TeV~\cite{ATLAS:2023kek}. Varying $m_{R_2}$ does not change our conclusion, which is that the constraints on Yukawa couplings deduced from experimental studies of $pp \to \tau \tau, \tau \nu$ (+ soft jets) at high-$p_T$'s are incompatible with those obtained from $R_{D^{(\ast )}}^\mathrm{exp}$. Obviously, that statement is valid to $2 \sigma$ and not to $3\sigma$ or more. It is therefore difficult to make a strong statement on this issue because the 
 uncertainties on the high energy end, related to the reconstruction of $\tau$ leptons can be questioned~\cite{Jaffredo:2021ymt}, and those on the low energy end related to the form factors used to compute $R_{D^*}$, may change once they are fully understood and the results of various lattice collaborations agree. Note, however, that the effect of propagation of $R_2$ has been properly taken into account. Notice also that dedicated experimental searches for a leptoquark signal in $pp \to \tau \tau$ could definitively exclude this scenario.

Since we have aligned the couplings with the down-quark mass, the tree level flavor changing neutral semileptonic processes $b \to s$ or $b \to d$ are forbidden. We thus cannot expect significant effects contributing $b \to s \nu \nu$ or $b \to s \ell \ell$ processes. The corresponding $b \to d$ rare transitions are further CKM suppressed.

\subsection{$\widetilde{R}_2$}
\label{sec:R2t}
Another interesting scenario that could potentially describe the deviation $\rddst^\mathrm{exp}>\rddst^\mathrm{SM}$ is the one with a doublet of $\widetilde R_2$ scalar  leptoquarks. In terms of the SM quantum numbers, $\widetilde R_2$ is specified by $(3,2)_{1/6}$, and it is peculiar because besides its coupling to a lepton doublet, it can also couple to a lepton singlet state, $N_R$, namely:
\begin{equation}
\label{eq:LR2tilde}
\mc{L}=-\widetilde{y}^{ij}_L \overline{d}^i\widetilde{R}_2^a \epsilon^{ab}L^{j,b}+\widetilde{y}^{iN}_R\overline{Q}^{i,a}\widetilde{R}_2^a N_R+\text{h.c.}\,.
\end{equation}
We again opt for a minimal setup and fix our model by choosing as non-zero Yukawa couplings:
\begin{equation}
  \label{eq:YukR2t}
  \widetilde{y}_L =
  \begin{pmatrix}
    0 & 0 & 0\\
    0 & 0 & 0\\
    0 & 0 & \widetilde{y}_L^{b\tau}
  \end{pmatrix}\,, \qquad \widetilde{y}_R = \begin{pmatrix}
     0\\
     \widetilde{y}_R^{sN} \\
     0
  \end{pmatrix} \,,
\end{equation}
needed to enhance $\rddst^\mathrm{SM}$
As before, the coupling to $N_R$ is in the down-quark basis. For the sake of simplicity we introduce only one massive sterile state $N_R$ but such that its mass is negligible with respect to all the other particles participating in $B\to D^\ast \tau N_R$. The interaction Lagrangian, in the mass basis of quarks, then reads:
\begin{equation}
\begin{aligned}
    \mc{L}=&-\widetilde{y}^{b\tau}_L (\overline{b}P_L \tau) \widetilde{R}_2^{2/3} +\widetilde{y}^{b\tau}_L (\overline{b}P_L \nu) \widetilde{R}_2^{-1/3}+\\
    &+\widetilde{y}^{sN}_R (\overline{s} P_R N_R) \widetilde{R}_2^{-1/3}+
    \widetilde{y}^{sN}_R V_{js}(\overline{u}_j P_R N_R) \widetilde{R}_2^{2/3} +\text{h.c.}\,,
\end{aligned}
\end{equation}
where the superscript in $\widetilde{R}_2^{Q}$ denotes the leptoquark's electric charge. It is important to emphasize that $N_R$ is not just a chirally flipped projection of the ordinary neutrino, but a completely different particle. As such, it entails the new physics contribution that does not interfere with the SM, nor with the BSM contribution involving left-handed neutrinos in the final state. In other words, the contribution involving $N_R$ always increases the decay width with respect to the SM. Schematically, the branching fraction of a decay mode
\begin{equation}
\mathcal{B}\,\propto \,\left|\mc{A}_{\text{SM}}+\mc{A}_{\mrm{NP}}^{\nu_L}\right|^2+\left|\mc{A}_{\text{NP}}^{N_R}\right|^2\,.
\end{equation}
In this model there are two tree level contributions to $\rddst$ which are described by two $N_R$-SMEFT coefficients~(\ref{eq:NRSMEFT}). It is a simple matter to read them off at the matching scale $\mu = m_{\widetilde R_2}$ and get:
\begin{equation}
    C^{(1)}_{\substack{Nldq \\ \phantom{N} \tau b s}}( m_{\widetilde R_2}) = 4\, C^{(3)}_{\substack{Nldq \\ \phantom{N} \tau b s}}( m_{\widetilde R_2}) = -\frac{1}{2} \  \widetilde y_R^{sN}\, \widetilde y_L^{b\tau*}  \,.
\end{equation}
Running from the matching scale down to $\mu = m_b$ is then made by means of Eq.~\eqref{eq:STtildematchandrun} so that one finally arrives at the low energy effective theory~\eqref{eq:LeffSL} with a scenario $\widetilde g_{S_R}(m_b) = 8.8\times \widetilde g_T(m_b)$.

\begin{figure}
    \centering
    \includegraphics[scale=0.40]{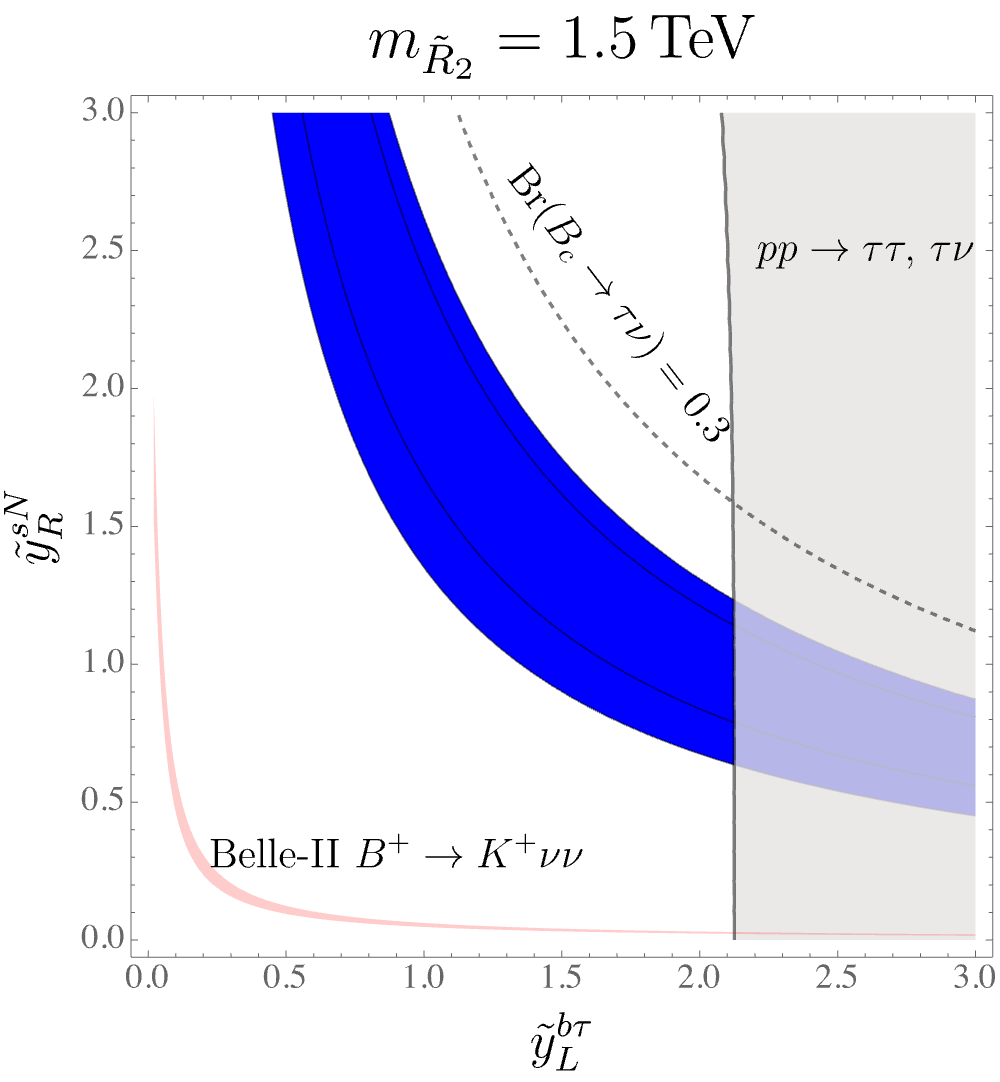}
    \caption{\small \sl $\widetilde{R}_2$ scenario at the high-energy scale $\mu=m_{\widetilde R_2}=1.5$~TeV. The blue band corresponds to the constraint arising from $\rddst^\mathrm{exp}$, while the exclusion from the high-$p_T$ tails corresponds to a shaded gray region. Limit on $\tau_{B_c}$ is enforced by the requirement, $
    \mathcal{B}(B_c\to \tau \nu)\leq 30\%$, is below the dashed curve in the plot. 
The red curves correspond to the recently measured $\mathcal{B}(B\to K \nu\nu)$~\cite{Belle-II:2023esi}.}    \label{fig:bounds_R2tilde} 
\end{figure}
Besides $\rddst$, one should keep the width of $B_c$ under control. This is usually enforced by requiring $\mathcal{B}(B_c\to \tau \nu)\leq 30\%$ which is indicated in Fig.~\ref{fig:bounds_R2tilde}. Constraints from $pp \to \tau\tau$ are obtained by the HighPT package in the leptoquark mediator mode. Conversely, $N_R$ coupled to $\tilde R_2$ is not included within the HighPT package. Thus, in order to quantify the agreement of $\tilde R_2$-mediated $pp \to N_R \tau$ with experimental data we first observe that in partonic processes $pp \to \tau E_{T,\mrm{miss}}$ there is no interference between the SM and the amplitude with $N_R$. Thus, the signature of a $\tilde R_2^{2/3}$-mediated process, e.g. $u_L \bar b_R \to \tau_L \bar N_R$ is experimentally indistinguishable from $R_2^{2/3}$-mediated process $u_R \bar b_L \to \tau_R \nu_{\tau}$, provided the couplings and the masses of the two LQs are equal. In this way, by carefully adjusting $R_2$ couplings, we can fully assess the agreement of $\tilde R_2$ couplings with $pp \to \tau \nu_\tau$ experimental searches. We show the combined $pp \to \tau \tau,\tau \nu$ constraints in Fig.~\ref{fig:bounds_R2tilde}.

More problematic, however, is the fact that this model generates a huge contribution to $\mathcal{B}(B\rightarrow K^{(*)}\nu\nu)$, so that accommodating the current $\rddst^\mathrm{exp}$ would result in $\mathcal{B}(B\rightarrow K^{(*)}\nu\nu)$ orders of magnitude larger than the current experimental bounds. For that reason this model should be discarded.

As a side remark, we can turn the above argument around and claim that 
this scenario can be used to describe the recently measured $\mathcal{B}(B\to K \nu\nu)$~\cite{Belle-II:2023esi}, $3\sigma$ larger than predicted in the SM, should the improved measurement of $\rddst^\mathrm{exp}$ lower the current average.

\subsection{$S_1$}
The scalar singlet, often referred to as $S_1$, is the last of the three possible scalar leptoquarks that can accommodate the experimental hint of LFUV, $\rddst^\mathrm{exp} >\rddst^\mathrm{SM}$, with a minimal number of Yukawa couplings. In terms of the SM quantum numbers this leptoquark is described by $(\bar 3,1)_{1/3}$. Its peculiarity is that it can couple to two fermion doublets or to two singlets, namely,
\begin{equation}
\label{eq:LagrS1}
    \mc{L}_{S_1} = y_L^{ij}\, \overline{Q^{C,a}_i} \,\epsilon^{ab}\, L_j^b\,S_1 +  y_R^{ij}\, \overline{u^C_i} e_j\,S_1 + \mathrm{h.c.}\,.
\end{equation}
For the minimal setup of Yukawa couplings we choose,
\begin{equation}
  \label{eq:S1Yuk}
  y_L =
  \begin{pmatrix}
    0 & 0 & 0\\
    0 & 0 & 0\\
    0 & 0 & y_L^{b\tau}
  \end{pmatrix}\,, \qquad y_R = \begin{pmatrix}
    0 & 0 & 0\\
    0 & 0 & y_R^{c\tau} \\
    0 & 0 & 0
  \end{pmatrix} \,,
\end{equation}
defined in the mass basis of the down-type quarks, as before. In its more explicit form the above Lagrangian reads:
\begin{equation}
\label{eq:LagrS1_2}
    \mc{L}_{S_1} = y_L^{b\tau}\, V^*_{ib}(\overline{u^C_i} P_L\, \tau)\,S_1 - y_L^{b\tau}\, (\overline{b^C} P_L\, \nu_{\tau})\,S_1 +y_R^{c\tau}\, (\overline{c^C} P_R \tau)\,S_1 + \mathrm{h.c.}\,.
\end{equation}
With the above choice of couplings one has, at the matching scale $\mu = m_{S_1} = 1.5~\e{TeV}$,
\begin{align}
\label{eq:S1treematching}
  C^{(1)}_{\substack{lequ\\ \tau\tau b c}}(m_{S_1}) & = -4\, C^{(3)}_{\substack{lequ\\ \tau\tau b c}}(m_{S_1}) = -\frac{y_L^{b\tau} {y_R^{c \tau}}^* }{2}\,,
\end{align}
i.e. the non-zero low energy effective couplings are $g_{S_L}$ and $g_T$, with \begin{equation}
g_{S_L}(m_{S_1})= -
\frac{v^2}{4 V_{cb}}\  
\frac{y_L^{b\tau} {y_R^{c \tau}}^* }{ m_{S_1}^2 }\,.
\end{equation}
Their relation at the matching scale $g_{S_L}=-4 g_T$, after running down to $\mu = m_b$~\eqref{eq:STmatchandrun}, becomes:
\begin{equation}\label{eq:gSTbis}
    g_{S_L}(m_b)\ = \ -8.8\times g_T(m_b)\,.
\end{equation}
In the $S_1$ case, and with the couplings chosen as in Eq.~\eqref{eq:S1Yuk}, one also gets a non-zero $g_{V_L}$, namely
\begin{equation}
g_{V_L}= 
\frac{v^2}{4 V_{cb}}\  
\frac{V_{cb} \, | y_L^{b\tau}|^2 }{ m_{S_1}^2 }\,.
\end{equation}

\begin{figure}[t]
    \centering
    \includegraphics[width=1.\textwidth]{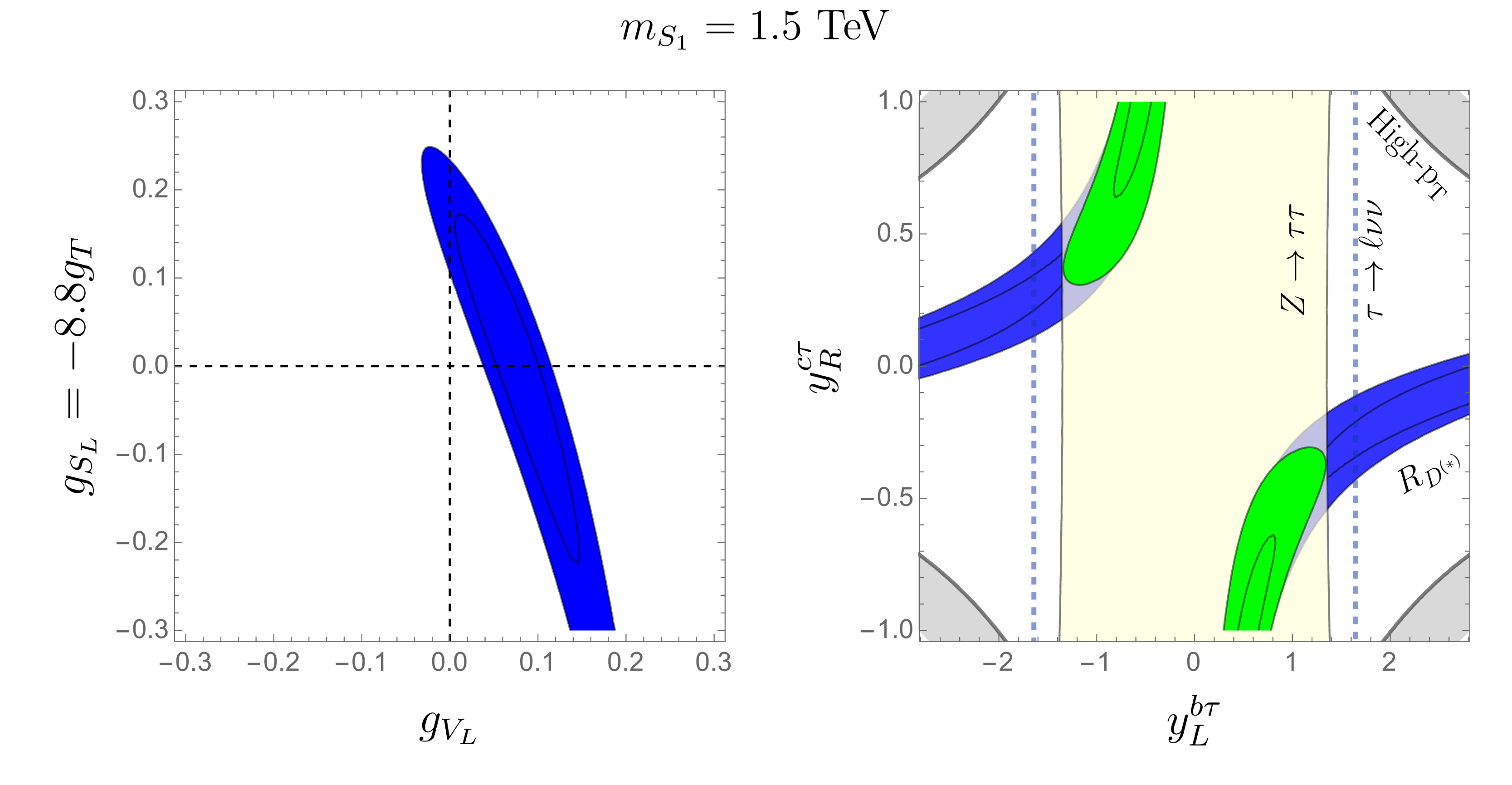}
    \caption{\small \sl In the left plot is shown the region of $g_{V_L}$ and $g_{S_L}(\mu=m_b) = -8.8 g_T(\mu=m_b)$ compatible with $R_{D^{(\ast)}}^\mathrm{exp}$ to $1\sigma$ and $2\sigma$. In the right plot are combined the constraints on the Yukawa couplings of the $S_1$ model specified in Eqs.~(\ref{eq:LagrS1},\ref{eq:S1Yuk}): Blue and yellow regions respectively depict the  $2\sigma$ consistency with $\rddst$ and $\mathcal{B}(Z\to \tau\tau)$. The latter is comparable with the constraint marked with dashed lines corresponding to the region allowed by $\mathcal{B}(\tau \to \mu \bar \nu_\mu\nu_\tau)$ to $2\sigma$. Note that in this case the gray regions are not allowed by the experimental studies of high-$p_T$ tails of $pp\to \tau\nu,\tau\tau$ (to $2\sigma$ as well). Green regions are the result of the global fit at $1$- and $2\sigma$ CL.}
    \label{fig:bounds_S1} 
\end{figure}

Contrary to the $R_2$ case, in this situation one can find a region in which all the constraints overlap for real values of couplings, which is shown in Fig.~\ref{fig:bounds_S1}. As before, the measured $\Gamma(Z\to \tau\tau)$ represents a powerful constraint on the Yukawa couplings but this time it is interesting to note that it is comparable to the constraint obtained from $\mathcal{B}(\tau \to \mu \bar \nu\nu )$ to which the leptoquark correction is also generated through a loop, cf. Ref.~\cite{Arnan:2019olv}. We have used Flavio package~\cite{Straub:2018kue,peter_stangl_2023_7994776} for the leptonic $\tau$ decays.

Clearly, this is the only acceptable single scalar leptoquark solution to the problem of $\rddst^\mathrm{exp} >\rddst^\mathrm{SM}$ involving a minimal number of parameters. 
Since this model is viable, we can use the region of allowed parameters shown in Fig.~\ref{fig:bounds_S1} and make several interesting predictions. 
\begin{itemize} 
\item[1.] In the previous Section we made sure that $\mathcal{B}(B_c\to \tau \nu)\leq 30\%$. In our $S_1$ model such a requirement is not necessary since the correction to $B_c\to \tau \nu$ is generated through $g_{S_L}$ and amounts to:
\begin{equation}
\frac{\mathcal{B}(B_c\to \tau \nu)^{S_1}}{\mathcal{B}(B_c\to \tau \nu)^\mathrm{SM}}  \in [1.13, 1.48]\,,\ \mathrm{where}\ 
\mathcal{B}(B_c\to \tau \nu)^\mathrm{SM} = (2.24 \pm 0.07)\% \times \left(\frac{V_{cb}}{0.0417}\right)^2,
\end{equation}
which in fact is a prediction of this model, and it is well below $\mathcal{B}(B_c\to \tau \nu)\leq 30\%$.
\item[2.] In the previous Section we also showed that the consistency with $\rddst^\mathrm{exp}$ resulted in a huge enhancement of $\mathcal{B}(B\to K\nu\nu)$. In our $S_1$ model this is not the case because there is no tree-level coupling to $s\nu$. It can however generate, through the box or penguin diagrams involving one $S_1$ and one $W$-boson, a contribution to $b \to s \tau\tau$ or $b \to s \nu_\tau \bar \nu_\tau$, as shown in Figs.~\ref{fig:S1Box},\ref{fig:S1Penguin}.
\begin{figure}
    \centering
    \begin{tabular}{lr}
         \includegraphics[scale=1.15]{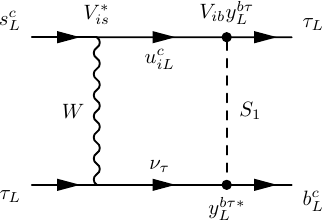} &  
    \end{tabular}
    \caption{\sl Dominant contribution to $b \to s \tau^+ \tau^-$ via box diagram.}
    \label{fig:S1Box}
\end{figure}
The leading contribution to $b \to s \tau \tau$ is due to virtual top quark in the box, as shown in Fig.~\ref{fig:S1Box} which then shifts the SM Wilson coefficients as
\begin{equation}
    \delta C_9 = -\delta C_{10} = -\frac{|y_L^{b\tau}|^2}{8\pi \alpha}\,\frac{x_t \log x_t}{x_t - 1} = -0.31\,|y_L^{b\tau}|^2\,, 
\end{equation}
where $x_q = m_q^2/m_{S_1}^2$. 
The contribution from the box diagram is computed in the broken electroweak phase with massive quarks so that 
GIM actually annuls all the $u_i$-mass independent terms, leading to a finite result (even in the unitary gauge) and to vanishing of the whole diagram if the quarks were mass degenerate. 
Comparing this to the SM Wilson coefficients, $C_9^\mrm{SM} = 4.2$, $C_{10}^\mrm{SM} = -4.1$, we use $\delta C_9/C_9^\mrm{SM} \approx \delta C_{10}/C_{10}^\mrm{SM} = -0.075|y_L^{b\tau}|^2$ and find:
\begin{equation}
\frac{\mathcal{B}(B_s\to \tau \tau)^{S_1}}{\mathcal{B}(B_s\to \tau \tau)^\mathrm{SM}}  \in [0.73, 0.98]\,,\quad \frac{\mathcal{B}(B\to K \tau \tau)^{S_1}}{\mathcal{B}(B\to K \tau \tau)^\mathrm{SM}}  \in [0.73, 0.98]\,\quad (@ 2 \sigma)\,.
\end{equation}
This is quite a remarkable result since the suppression occurs only due to $y_L^{b\tau}\neq 0$. That means that even if $R_{D^{(\ast )}}^\mathrm{exp}$ were equal to $R_{D^{(\ast )}}^\mathrm{SM}$, one could simply have $y_R^{c\tau}\to 0$, and still have the above suppression of the $b\to s\tau\tau$ rates. Knowing that $y_L^{b\tau}$ is very difficult to constrain either through the LHC studies of high-$p_T$ tails of $pp\to \tau\tau$, or via the low energy constraints, such as $\mathcal{B}(\Upsilon(nS) \to \tau\tau)$, actually measuring $\mathcal{B}(B_s\to \tau \tau)$ and/or $\mathcal{B}(B\to K^{(\ast )} \tau \tau)$ would be the only way to understand whether or not the above suppression indeed takes place. 
\item[3.] 
\begin{figure}[!h]
    \begin{tabular}{lcccr}
           \includegraphics[scale=1.15]{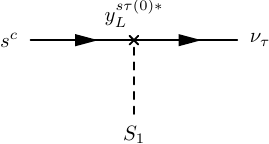} &&&
           \includegraphics[scale=1.15]{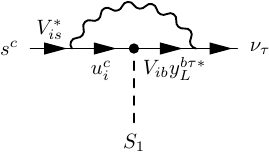}
    \end{tabular}
    \caption{\sl Dominant contribution to the $y_L^{s\tau}$ renormalized vertex leading to $b \to s \nu_\tau \bar \nu_\tau$.}
    \label{fig:S1Penguin}
\end{figure}
Penguin contribution to $b \to s \nu_\tau \bar \nu_\tau$: 
We already stated that the models we consider in this paper are not meant to solve the $2.7\sigma$ discrepancy between the first measurement of $\mathcal{B}(B\to K\nu\nu)^\mathrm{Belle~II}$ and its SM prediction. Since in this case there is no tree-level contribution to $b\to s\nu\nu$ there is no worry that $\mathcal{B}(B\to K^{(\ast )}\nu\nu)$ in this model could then be in conflict with experimental upper bounds. This model in fact leads to a loop induced contribution to $b \to s \nu_\tau\nu_\tau$ which is again proportional to $|y_L^{b\tau}|^2$. Main features of the computation of the diagrams shown in Fig.~\ref{fig:S1Penguin} are described in Appendix~\ref{sec:B}. In terms of the relevant Wilson coefficient, 
\begin{equation}
C_L^{ij}= C_L^\mathrm{SM} \delta^{ij} + \delta C_L^\mathrm{S_1}\,\delta^{i\tau} \delta^{j\tau}\,,\quad  \delta C_L^\mathrm{S_1}=  \frac{|y_L^{b\tau}|^2}{16 \pi \alpha} \, \sum_{i=u,c,t} \frac{\lambda_i}{\lambda_t} \,x_W \,g(x_i,x_W)\,,
\end{equation}
where $C_L^\mathrm{SM}=-6.32(7)$~\cite{Brod:2010hi}. In the $S_1$ contribution we sum over all up-quarks in the loop that contribute to the renormalized $y^L_{s\tau}$. Besides $x_i = m_{u_i}^2/m_{S_1}^2$, we also introduced $x_W = m_W^2/m_{S_1}^2$, and the CKM factors $\lambda_i = V_{ib} V_{is}^*$. The explicit expression for the loop function $g(x_i,x_W)$ can be found in Eq.~\eqref{eq:gLoop}. Here we just note an important feature that a mild dependence of the loop function on $x_i$
leads to an efficient GIM cancellation. For illustration, we observe that $g(x_u,x_W) \approx g(x_c,x_W) = 31.0 - 36.8\,i$ is not much different from $g(x_t,x_W) = 19.1 - 36.2\,i$. The imaginary part arises from the fact that all fermions propagating in the loop can be on their mass shell. 
Finally the $S_1$ contribution to the Wilson coefficient is
\begin{equation}
    C_L^\mathrm{S_1}=  (-9.3 + 0.4\,i)\times 10^{-2}\, |y_L^{b\tau}|^2\,.
\end{equation}
The resulting shift of the physical decay rates is:
\begin{equation}
\frac{\mathcal{B}(B\to K^{(\ast )} \nu \nu)^{S_1}}{\mathcal{B}(B\to K^{(\ast )}\nu \nu)^\mathrm{SM}} = \left| 1 + \frac{\delta C_L^\mathrm{S_1}}{3\,C_L^\mathrm{SM}} \right|^2 \in [1.001, 1.02]\, \quad(@2\sigma)\,.
\end{equation}

It is important to note that this prediction depends heavily on the assumption that $y_L^{s\tau}(\mu)=0$ (eq. \ref{eq:S1Yuk}). In principle, that does not have to hold since the current best bound is $y_L^{s\tau}\lesssim 1.5$ due to high-$p_T$ constraints and $D_s\rightarrow \tau \nu$ decay. Such large values for $y_L^{s\tau}$ would drastically change the prediction since it causes a tree-level effect in the given process.

\item[4.] Among other (semi-)leptonic decays, we may expect an appreciable contribution to $t \to b \tau \nu$ decay. Even though it is proportional to $|y_L^{b\tau}|^2$ that contribution is too small to be distinguished experimentally.\footnote{The latest reported $\mathcal{B}(t\to b\tau\nu)=0.105(1)(7)$~\cite{CMS:2019snc} is measured with $7\%$ of systematic uncertainty, which is much larger than the leptoquark contribution to this decay.}
\item[5.] A term proportional to $V_{ub} |y_L^{b\tau}|^2$ can contribute to the $b \to u \tau \nu$ amplitude and thus it can modify $\mathcal{B}(B^- \to \tau \nu)$ and $\mathcal{B}(B \to \pi \tau \nu)$. The leading term interferes with SM and gives at most $3\%$ enhancement of the SM branching fractions.
\item[6.] Besides $\mathcal{B}(B\to D^{(\ast )}\tau \nu)$, one can infer a number of observables from the angular distribution of this decay, cf. for example Refs.~\cite{Sakaki:2013bfa,Becirevic:2016hea,Blanke:2018yud,Blanke:2019qrx,Mandal:2020htr,Bobeth:2021lya}. In the case of $\tau$ in the final state, the fraction of the decay rate to a longitudinally polarized $D^\star$ has been measured and the two measurements do not agree, $F_L^{D^*,\mrm{Belle}} = 0.60(9)$ and $F_L^{D^*,\mrm{LHCb}} = 0.43(7)$, former being larger and latter consistent with the SM prediction, $F_L^{D^*,\mrm{SM}} = 0.46(1)$. Belle also managed to measure the $\tau$-polarization asymmetry in the $B\to D^\ast \tau\nu$, and found $P_\tau^{D^*,\mrm{Belle}} = -0.37\pm 0.54$ which, with increasing accuracy, may become an important observable to select among various BSM scenarios. Its SM value is known, $P_\tau^{D^*,\mrm{SM}} = -0.51(2)$. In our $S_1$, we obtain:
\begin{equation}
F_L^{D^*} =  0.44(1)\,,\qquad P_\tau^{D^*} = -0.53(3)\,.
\end{equation}
Similarly, for the forward backward asymmetry we find
\begin{equation}
A_\mathrm{fb}^{D^*} =  -0.05(1)\,,\qquad A_\mathrm{fb}^{D} =   0.33(1)\,,
\end{equation}
the values that are to be compared to the SM predictions $A_\mathrm{fb}^{D,\mrm{SM}} =   0.3600(4)$, and 
$A_\mathrm{fb}^{D^\ast ,\mrm{SM}} =   -0.06(1)$.
\end{itemize}
\begin{figure}
    \centering
    \includegraphics[width=0.9\textwidth]{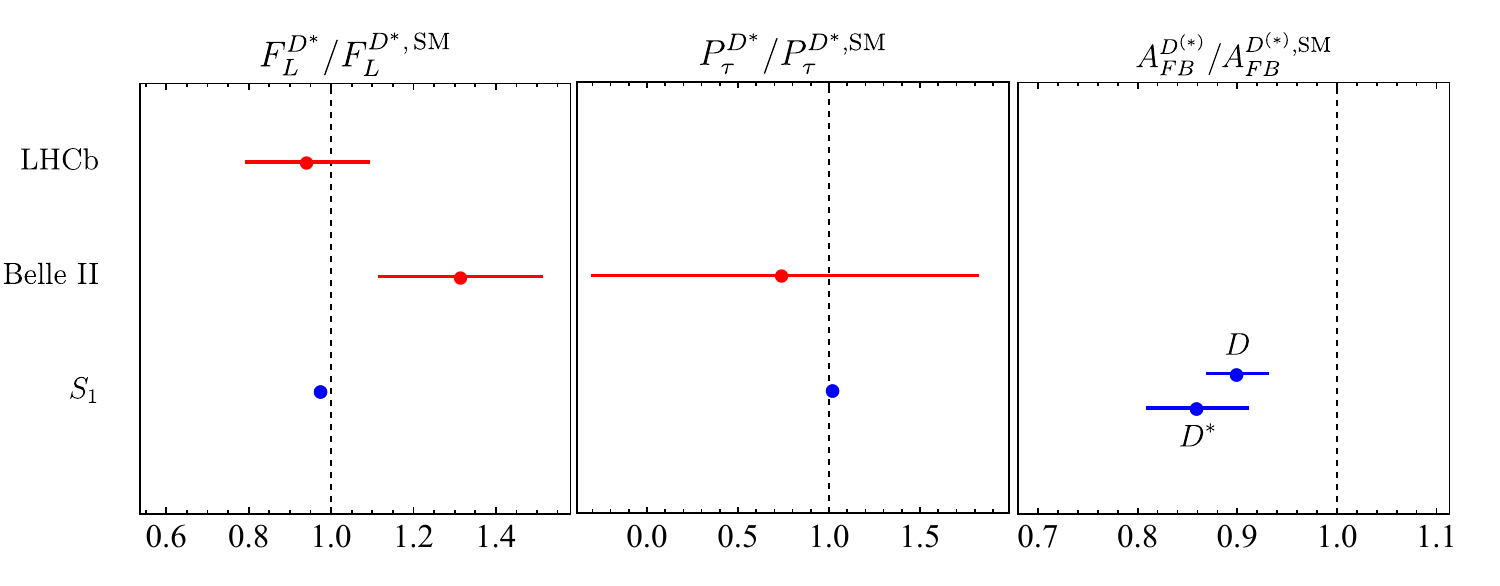}
    \caption{\small \sl Predictions of our $S_1$ model for three observables relevant to $B\rightarrow D^{(*)} \tau \nu$, namely (i) $F_L^{D^\ast}$, the fraction of longitudinally polarized $D^\ast$, (ii) $P_\tau^{D^\ast}$, the $\tau$-lepton polarization asymmetry, and (iii) $A_\mathrm{fb}^{D^{(\ast )}}$, the integrated forward-backward asymmetry. Shown are the ratios of our predictions with respect to the SM values. Dashed vertical lines thus correspond to the SM. We also display the experimental results when available. }
    \label{fig:polarization_S1} 
\end{figure}

\subsection{$S_1$ alternative}
\label{sec:S1alt}

\begin{figure}
    \centering
    \includegraphics[scale=0.65]{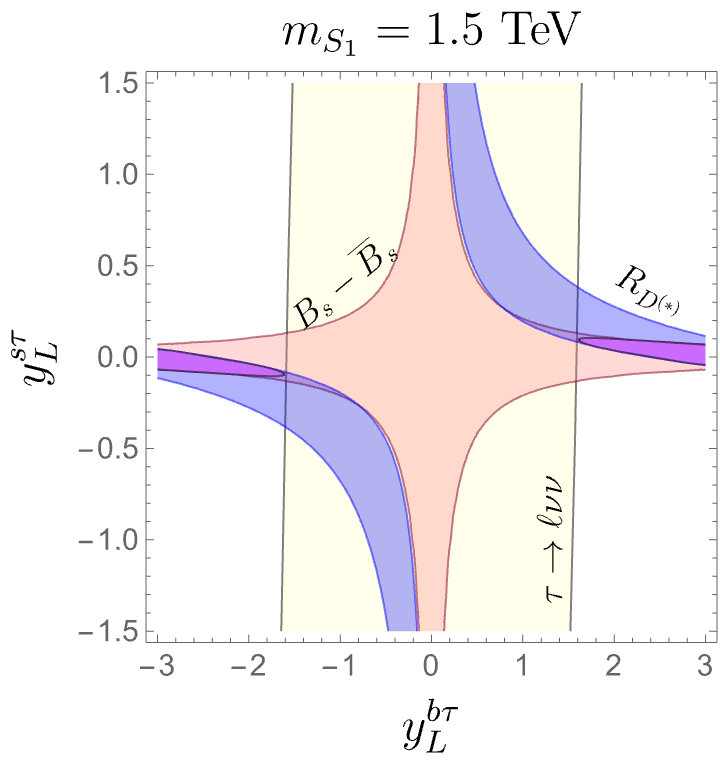}
    \caption{\small \sl $S_1$ with left-handed couplings only. Consistency with experimental information to $2 \sigma$ results in the purple bounds obtained by combining $\Delta m_{B_s}$ and $\rddst$ constraints, and the light yellow region allowed by $\mathcal{B}(\tau \to \ell \nu\bar \nu)$ with $l = e, \mu$. Clearly the two regions do not overlap, thus making this model a nonviable solution to $\rddst^\mathrm{exp} >\rddst^\mathrm{SM}$.}
    \label{fig:bounds_S1V} 
\end{figure}

In Eq.~\eqref{eq:S1Yuk} we chose both the left- and the right-handed Yukawa couplings. In that way we provided a viable solution to $\rddst^\mathrm{exp} >\rddst^\mathrm{SM}$, as shown in Fig.~\ref{fig:bounds_S1}. However, as it can be seen in Lagrangian~\eqref{eq:LagrS1_2}, one could also opt for left-handed Yukawa couplings only and generate a contribution to $\rddst$. One (minimal) possibility is to choose
\begin{equation}
  \label{eq:S1Yuk_VA}
  y_L =
  \begin{pmatrix}
    0 & 0 & 0\\
    0 & 0 & y_L^{s\tau}\\
    0 & 0 & y_L^{b\tau}
  \end{pmatrix}\,, \qquad y_R = 0
  \,,
\end{equation}
so that the only non-zero coupling in LEET~\eqref{eq:LeffSL} is $g_{V_L}$, which is expressed by the first term in Eq.~\eqref{eq:Vmatch}, i.e.
\begin{equation}
g_{V_L}=- \frac{v^2}{m_{S_1}^2} \frac{V_{cs}}{V_{cb}} y_L^{s\tau *}y_L^{b\tau}.
  \end{equation}
It is the CKM enhancement that makes this model appealing but it nevertheless gets excluded by the $B_s-\overline B_s$ mixing (i.e. $\Delta m_{B_s}$) which is incompatible with the constraint stemming from $\mathcal{B}(\tau \to \mu \nu\bar \nu)$, as it can be seen in Fig.~\ref{fig:bounds_S1V}. Note that in the numerical analysis we used the lattice QCD result for the hadronic parameters $f_{B_s} \sqrt{\hat B_{B_s}}=256(6)$~MeV, as obtained from the lattice QCD simulation with $N_f=2+1+1$ sea quark flavors~\cite{Dowdall:2019bea}. The feature of this model that we emphasized above, namely that the constraints arising from $\Delta m_{B_s}^\mathrm{exp}$ and from $\mathcal{B}(\tau \to \mu \nu\bar \nu)^\mathrm{exp}$ are not mutually compatible, would be even more pronounced if we used the world average lattice QCD result with $N_f=2+1$~\cite{FlavourLatticeAveragingGroupFLAG:2021npn}, $f_{B_s} \sqrt{\hat B_{B_s}}=274(8)$~MeV.

\section{Summary}

In this paper we revisited the possibilities of explaining the experimental hint of LFUV suggesting that $\rddst^\mathrm{exp} >\rddst^\mathrm{SM}$ to more than $3\sigma$, by extending the SM by a single scalar leptoquark in the minimal setup, i.e. with the minimal number of Yukawa couplings. Of three scenarios that could potentially contribute to $\rddst$ we found that only $S_1$ leptoquark can provide the desired enhancement without being in conflict with other constraints, most notably those stemming from $\mathcal{B}(Z\to \tau\tau)^\mathrm{exp}$ and from the LHC studies of the tails of differential cross section of $pp\to \tau\tau, \tau\nu$ (+ soft jets) at high $p_T$. 
While this latter constraint will be steadily improved with higher luminosity of the experimental data, it is important to emphasize that an important progress should be made on the low energy side too. Current lattice QCD estimates of the hadronic matrix elements relevant to $B\to D^\ast \ell \nu$ still suffer from important systematic uncertainties: the shapes of the form factors are not clear and, apart from the dominant form factor $A_1(q^2)$ various collaborations do not agree among themselves. In such a situation, we decided to use the form factors extracted from the experimental studies regarding the angular distribution of $B\to D^\ast (\to D\pi) l\nu$, with $l=(e,\mu)$, and to evaluate $R_{D^\ast}^\mathrm{SM}$ we needed only one information from lattice QCD, namely 
[$A_0(q^2_\mathrm{max})$]. This is also licit in the rest of our considerations since our main assumption is that the new physics (leptoquarks) can couple to the third generation of leptons only.\footnote{Note also that the tensor form factors were recently computed on the lattice~\cite{Harrison:2023dzh}, which we use in form of the ratios $T_{1,2,3}(q^2)/A_1(q^2)$. } 

Our only scenario, that we deem as viable, is the one with $S_1$ leptoquark and with Yukawa couplings to both left- and right-handed quark/lepton doublets. We checked that the other possibility, in which only the left-handed couplings are allowed to be non-zero, is not simultaneously consistent with the constraints arising from $\Delta m_{B_s}$ and from $\mathcal{B}(\tau \to \ell \nu\bar \nu)$. We provided several predictions that can help supporting or invalidating the model we propose. 

As for the other models: $R_2$-model exhibits strong tension between the constraints arising from $\rddst$ and from the high-$p_T$ tails mentioned above. That tension disappears, however, at $3\sigma$ and even though we deem the model unsatisfactory, one should monitor how the world average of $\rddst^\mathrm{exp}$ will evolve and in what way will move the constraints coming from experimental studies of high-$p_T$ tails with the next acquisition of data at the LHC.

On the other hand, the $\widetilde R_2$ model cannot provide an explanation of $\rddst^\mathrm{exp} >\rddst^\mathrm{SM}$ because it necessarily results in a huge contribution to $b\to s\nu\nu$ which then overshoots the experimental bounds on $\mathcal{B}(B\to K^{(\ast )} \nu\nu)$ by orders of magnitude. 

\vspace*{2cm}

\section*{Acknowledgments}
D.B. received a support of the European Union’s Horizon Europe research and innovation programme under the Marie Skłodowska-Curie Staff Exchange agreement No~101086085 - ASYMMETRY and -Curie grant agreement No~860881-HIDDeN.
S.F., N.K. and L.P. acknowledge financial support from the Slovenian Research Agency (research core funding No. P1-0035 and N1-0321). N.K. acknowledges support of the visitor programme at CERN Department of Theoretical Physics, where part of this work was done.

\appendix

\newpage
\section*{Appendix}

\section{$B_s$ mixing}
The $S_1$ leptoquark with the choice of couplings as in Sec.~\ref{sec:S1alt} contributes via a box-diagram to the $B_s$-$\overline{B_s}$ mixing amplitude. For convenience we match directly to the low-energy effective theory  and account for the QCD renormalization group running between the scales $\mu = m_{S_1}$ and $\mu = m_b$. The relevant Lagrangian reads:
\begin{equation}
  \label{eq:LeffBsMix}
  \mathcal{L}_{bs} = -\frac{4 G_F}{\sqrt{2}}\lambda_t^2 C^{LL}_{bs} (\bar s_L
  \gamma^\mu b_L) (\bar s_L
  \gamma_\mu b_L) + \mrm{h.c.}\,,
\end{equation}
in the notation in which $\lambda_t = V_{tb}^* V_{ts}$. 
At the matching scale, $\mu=m_{S_1}=1.5$~TeV, for the Wilson coefficient we get
\begin{equation}
  \label{eq:CLLbs}
  C^{LL(S_1)}_{bs}(m_{S_1}) = \frac{v^2}{256 \pi^2 m_{S_1}^2} \frac{\left(y_L^{s\tau*} y_L^{b\tau}\right)^2}{\lambda_t^2} \,,
\end{equation}
which, due to 2-loop running down to $\mu = m_b$, is rescaled by a factor $\eta_{S_1}=0.617$, i.e.
\begin{equation}
    C^{LL(S_1)}_{bs}(m_b) = \eta_{S_1}  C^{LL(S_1)}_{bs}(m_{S_1})\,.
\end{equation}
This is to be added to the SM contribution to the same Wilson coefficient~\cite{Buras:1990fn,Artuso:2015swg}:
\begin{equation}
C_{b s}^{L L(\mathrm{SM})}\left(m_b\right)  =\eta_B \frac{m_W^2 S_0\left(m_t^2 / m_W^2\right)}{16 \pi^2 v^2} 
 =0.862(3) \times 10^{-3}\,,
\end{equation}
where $\eta_B = 0.55$.\footnote{Note that we use the renormalization group invariant definitions of $\eta_{S_1}$, $\eta_B$ and of bag parameter $\hat B_{B_s}$.} Finally, the frequency of oscillations of the $B_s$-$\overline{B_s}$ system is
\begin{equation}
\label{eq:DeltaMs}
\Delta M_{B_s} =\Delta M_{B_s}^\mrm{SM}\ \left|1+\frac{C^{LL(S_1)}_{bs}(m_b)}{C^{LL\mrm{(SM)}}_{bs}(m_b)}\right|\,,
\end{equation}
where for the SM value, after using $f_{B_s} \sqrt{\hat B_{B_s}} = 0.256(6)$~GeV~\cite{Dowdall:2019bea}, we obtain
\begin{equation}
	\Delta M_{B_s}^\mrm{SM} = \frac{4 m_{B_s}}{3}\,f_{B_s}^2 \hat B_{B_s} \,|\lambda_t|^2\, \frac{C^{LL\mrm{(SM)}}_{bs}(m_b)}{v^2} = 17.1(8)~\mathrm{ps}^{-1} \left(\frac{\lambda_t}{0.04106}\right)^2\,.
\end{equation}
Precise prediction depends on the values taken for the bag parameters and CKM elements, which is to be compared to 
$\Delta M_{B_s}^\mrm{exp}=17.765(6)~\mathrm{ps}^{-1}$~\cite{ParticleDataGroup:2022pth}. Notice also that the above SM value of $\Delta M_{B_s}$ does not include the uncertainty in $\lambda_t$ which is actually significant, cf. Ref.~\cite{Becirevic:2023aov}. 
In our numerical analysis we used the above-mentioned values, including the only lattice QCD estimate of $f_{B_s} \sqrt{\hat B_{B_s}}$ obtained from simulations with $N_f=2+1+1$ flavors. If, instead, we used the world average value, $f_{B_s} \sqrt{\hat B_{B_s}} =0.274(8)$~GeV, as obtained from simulations with $N_f=2+1$ flavors~\cite{FlavourLatticeAveragingGroupFLAG:2021npn}, we would have, $\Delta M_{B_s}^\mrm{SM}= (19.6\pm 1.1)~\mathrm{ps}^{-1}$, modulo uncertainty on $\lambda_t^2$.

\section{Penguin contribution to $b\to s\nu\nu$ in our $S_1$ model \label{sec:B}}
In this Appendix we describe the weak-interaction renormalization of $y_L^{s\tau}$, a coupling that is assumed to vanish at tree-level. In the process $S_1^* \to s \nu_\tau$ presented in Fig.~\ref{fig:S1Penguin} we choose the kinematical point such that $S_1$ has an arbitrary time-like momentum $q$ while the fermions are massless and on-shell. We calculate the right diagram $i\Delta y_{L}^{s\tau *}(q^2)$ in Fig.~\ref{fig:S1Penguin} in $R_\xi$ gauge and find it gauge-dependent.\footnote{Similar 1-particle-irreducible Goldstone-mediated diagrams are proportional to $m_\tau^2$ and we accordingly neglect them.} Clearly, the amplitude also depends on momentum $q^2$. The renormalized coupling $y_L^{s\tau}(q^2) = y_{L}^{s\tau(0)} + \Delta y_L^{s\tau}(q^2)$ is the sum of the loop- and tree-level contribution $y^{s\tau(0)}_{L}$. We fix the latter by the on-shell renormalization condition:
\begin{equation}
\label{eq:RenCond}
    y_L^{s\tau}(q^2 = m_{S_1}^2) = 0\,.
\end{equation}
Notice that once the weak radiative corrections are present, the $S_1$ Yukawa couplings run with $q^2$ and we can only set $y_L^{s\tau} =0$ at a given momentum scale which we choose to be such that the on-shell amplitude $S_1^* \to s \nu_\tau$ vanishes. In $\Delta y_{L}^{s\tau}(q^2)$ we have not considered 1-particle-reducible diagrams with $u_i \leftrightarrow s$ flavor-changing loops on the quark leg since these are $q^2$-independent and are thus removed by the renormalization condition~\eqref{eq:RenCond}.

The resulting $y_L^{s\tau}(q^2)$ is finite and gauge-independent:
\begin{equation}
    y_{L}^{s\tau *}(q^2) =\frac{g^2\,  y_{L}^{b\tau *}}{32 \pi^2} \,\sum_{i = u,c,t} \lambda_i\,\frac{m_{S_1}^2}{q^2}\,h\left(q^2/m_{S_1}^2, x_i, x_W \right)\,,
    \end{equation}  
where the loop function reads
    \begin{equation}
\begin{split}
    h(z,x_q,x_W) &=   -2 (x_q-1) z\, \Li \left(1-\frac{x_W}{x_q-1}\right)-
    2 x_q (z-1) \,\Li \left(1-\frac{x_q}{x_W}\right)\\
   &\quad +2 (x_q-z)\,
    \Li\left(1-\frac{x_W}{x_q-z}\right)-z(x_q-1) \log^2\left(\frac{x_W}{x_q-1}\right)\\
    &\quad +(x_q-z) \log^2\left(\frac{x_W}{x_q-z}\right)\,.
\end{split}
\end{equation}
We have expressed the loop diagram in terms of Passarino-Veltman functions~\cite{tHooft:1978jhc} with the help of FeynCalc package~\cite{Shtabovenko:2023idz}. To obtain analytic expressions for certain $C_0$ functions we have used Package-X~\cite{Patel:2016fam} and the results were numerically cross-checked against numerical values returned by program LoopTools~\cite{Hahn:1998yk,Hahn:2010zi}. At the scale of $B$ meson the momentum transfer $q^2$ is well below  $m_{S_1}^2$ and we can safely take a limit $z \to 0$, which leads us to
\begin{equation}
\begin{split}
    \label{eq:gLoop}
        g(x_i, x_W) &\equiv \lim_{z\to 0} \frac{h(z,x_i,x_W)}{z} \\
        &=-2 x_i\,\Li\left(1-\frac{x_q}{x_W}\right)-2 \,\Li\left(1-\frac{x_W}{x_i}\right) +\frac{2 x_i \log \left(\frac{x_W}{x_i}\right)}{x_i-x_W}\\
        &\quad-(x_i-1) \left[2 \,\Li\left(1-\frac{x_W}{x_i-1}\right)+\log^2\left(\frac{x_W}{x_i-1}\right)\right]-\log^2\left(\frac{x_W}{x_i}\right)\,.
\end{split}
\end{equation}

\bibliographystyle{h-physrev}
\bibliography{draft}

\end{document}